\documentclass[aps,pra,10pt,reprint,superscriptaddress,twocolumn,showpacs,showkeys,a4paper,longbibliography]{revtex4-1}

\usepackage{amsfonts,amsmath,amssymb}
\usepackage{bbm,bm}
\usepackage[squaren]{SIunits}
\usepackage[usenames,dvipsnames]{color}
\usepackage[pdftex]{graphicx}
\usepackage{hyperref}

\usepackage[normalem]{ulem}
\usepackage{slashed}

\definecolor{mygreen}{rgb}{0,0.5,0} 
\definecolor{myblue}{rgb}{0,0,0.75} 
\definecolor{myyellow}{rgb}{0.87,0.8,0.47} 
\definecolor{mymagenta}{cmyk}{0,1,0,0.12}

\newcommand{\var}{\mathrm{var}}
\newcommand{\cov}{\mathrm{cov}}

\newcommand{\ave}[1]{\ensuremath{\langle#1\rangle}}
\newcommand{\norm}[1]{\ensuremath{\lvert#1\rvert}}

\newcommand{\NA}{N_{\rm A}}
\newcommand{\NL}{N_{\rm L}}

\newcommand{\myf}{\hat{f}}
\newcommand{\myJ}{\hat{J}}
\newcommand{\myj}{\hat{\jmath}}
\newcommand{\myS}{\hat{S}}

\newcommand{\myH}{\hat{H}}
\newcommand{\myK}{\hat{K}}

\newcommand{\Jx}{\myJ_x}
\newcommand{\Jy}{\myJ_y}
\newcommand{\Jxy}{\myJ_{x,y}}
\newcommand{\jx}{\myj_x}
\newcommand{\jy}{\myj_y}
\newcommand{\jxy}{\myj^{(i)}_{x,y}}

\newcommand{\mJx}{J_x}
\newcommand{\mJy}{J_y}
\newcommand{\mFz}{J_z}

\newcommand{\Ji}{\myJ_i}

\newcommand{\Fz}{\myJ_z}

\newcommand{\fz}{\myf_z}

\newcommand{\K}{\myK_{\theta}}

\newcommand{\Sx}{\myS_x}
\newcommand{\Sy}{\myS_y}
\newcommand{\Sz}{\myS_z}

\newcommand{\Si}{\myS_i}

\newcommand{\mSx}{S_x}
\newcommand{\mSy}{S_y}

\newcommand{\supin}{^{({\rm in})}}
\newcommand{\supout}{^{({\rm out})}}
\newcommand{\supplus}{^{(+)}}
\newcommand{\supminus}{^{(+)}}

\newcommand{\dd}{D_2}
\newcommand{\kA}{\kappa_1}
\newcommand{\kB}{\kappa_2}
\newcommand{\kAB}{\kappa_{1,2}}

\newcommand{\HH}{\myH_{\rm eff}}

\newcommand{\Bz}{B_z}

\newcommand{\phN}{\Phi_{\rm RO}}

\newcommand{\phiL}{\Phi_{\rm \LIN}}
\newcommand{\phiNL}{\Phi_{\rm AOC}}

\newcommand{\unktwo}{{\cal Y}}
\newcommand{\cpl}{{\kappa}}

\newcommand{\LIN}{LTE}

\newcommand{\tS}{\tilde{S}}
\newcommand{\btS}{\tilde{\bf S}}
\newcommand{\tSz}{\tS_z}
\newcommand{\tSy}{\tS_y}
\newcommand{\tSx}{\tS_x}

\newcommand{\EN}{{\rm EN}}
\newcommand{\OD}{{\rm OD}}

\begin{document}

\title{Ultrasensitive Atomic Spin Measurements with a Nonlinear Interferometer}

\newcommand{\ICFOAddress}{ICFO-Institut de Ciencies Fotoniques, Mediterranean Technology Park, 08860 Castelldefels (Barcelona), Spain}
\newcommand{\ICREAAddress}{ICREA - Instituci\'{o} Catalana de Re{c}erca i Estudis Avan\c{c}ats, 08015 Barcelona, Spain}

\author{R.~J.~Sewell}
\email[]{robert.sewell@icfo.es}
\affiliation{\ICFOAddress}

\author{M.~Napolitano}
\affiliation{\ICFOAddress}

\author{N.~Behbood}
\affiliation{\ICFOAddress}

\author{G.~Colangelo}
\affiliation{\ICFOAddress}

\author{F.~Martin Ciurana}
\affiliation{\ICFOAddress}

\author{M.~W.~Mitchell}
\affiliation{\ICFOAddress}
\affiliation{\ICREAAddress}


\date{\today}

\begin{abstract}
We study nonlinear interferometry applied to a measurement of atomic spin and demonstrate a sensitivity that cannot be achieved by any linear-optical measurement with the same experimental resources.  
We use alignment-to-orientation conversion, a nonlinear-optical technique from optical magnetometry, to perform a nondestructive measurement of the spin alignment of a cold $^{87}$Rb atomic ensemble.  
We observe state-of-the-art spin sensitivity in a single-pass measurement, in good agreement with covariance-matrix theory.  
Taking the degree of measurement-induced spin squeezing as a figure of merit, we find that the nonlinear technique's experimental performance surpasses the theoretical performance of any linear-optical measurement on the same system, including optimization of probe strength and tuning.  
The results confirm the central prediction of nonlinear metrology, that superior scaling can lead to superior absolute sensitivity.
\end{abstract}

\pacs{42.50.Ct,42.50.Lc,03.67.Bg,07.55.Ge}
\keywords{quantum metrology, quantum optics, optical magnetometer, spin squeezing,entanglement}

\maketitle


\section{Introduction}
Many sensitive instruments naturally operate in nonlinear regimes.
These instruments include optical magnetometers employing spin-exchange relaxation-free~\cite{Allred2002} and nonlinear~\cite{Budker2000a} magneto-optic rotation and interferometers employing Bose-Einstein condensates~\cite{Schumm2005a,Jo2007b,Jo2007c,Baumgaertner2010a}.
State-of-the-art magnetometers~\cite{Wolfgramm2010,Napolitano2011,Horrom2012,Hamley2012,Sewell2012,WolfgrammNPhot2013} and interferometers~\cite{Esteve2008,Riedel2010,Gross2010,Gross2011a,Luecke2011,Bucker2011a,Brahms2011,Berrada2013a} are quantum-noise limited and have been enhanced using techniques from quantum metrology~\cite{Caves1981,Giovannetti2004,Giovannetti2006,Giovannetti2011}.

A nonlinear interferometer experiences phase shifts $\phi$ that depend on $N$, the particle number, e.g. $\phi = \cpl N \unktwo$ for a Kerr-type nonlinearity $\unktwo$, where $\cpl$ is a coupling constant. 
This number-dependent phase implies a sensitivity $\Delta \unktwo \ge  (\cpl N)^{-1} \Delta \phi$, and if the nonlinear mechanism does not add noise beyond the $\Delta \phi = N^{-1/2}$ shot noise, the sensitivity  $\Delta \unktwo \propto N^{-3/2}$  even without quantum enhancement.  
Such a nonlinear system was identified in theory by Boixo {\em et al.}~\cite{Boixo2008a} and implemented with good agreement by Napolitano {\em et al.}~\cite{NapolitanoNJP2010,Napolitano2011}.
In contrast, entanglement-enhanced linear measurement achieves at best the so-called ``Heisenberg limit'' $\Delta \phi = N^{-1}$.  
The faster scaling of the nonlinear measurement suggests a decisive technological advantage for sufficiently large $N$~\cite{Luis2004,Luis2007,Rey2007a,Roy2008,Choi2008,Woolley2008,Boixo2008a,Boixo2009,Chase2009,Tacla2010,Tiesinga2013a}.  
On the other hand, no experiment has yet employed improved scaling to give superior absolute sensitivity, and several theoretical works~\cite{JavanainenPRA2012,Zwierz2010,Demkowicz-Dobrzanski2012,Hall2012a} cast doubt upon this possibility for practical and/or fundamental reasons.  

Here, we demonstrate that a quantum-noise-limited nonlinear measurement can indeed achieve a sensitivity unreachable by any linear measurement with the same experimental resources. 
We use nonlinear Faraday rotation by alignment-to-orientation conversion (AOC) \cite{Budker2000a}, a practical magnetometry technique~\cite{Budker2000a}, to make a nondestructive measurement of the spin alignment of a sample of $^{87}$Rb atoms~\cite{Koschorreck2010b,Sewell2012}.
AOC measurement employs an optically-nonlinear polarization interferometer, in which the rotation signal is linear in an atomic variable but nonlinear in the number of photons.
We have recently used AOC to generate spin squeezing by quantum nondemolition measurement~\cite{SewellNatPhot2013}, resulting in the first spin-squeezing-enhanced magnetometer ~\cite{Sewell2012}.  Here we show that this state-of-the-art sensitivity results from the nonlinear nature of the measurement, and could not be achieved with a linear measurement.
We demonstrate a scaling $\Delta \mJy \propto \NL^{-3/2}$, where $\NL$ is the photon number and $\mJy$ is an atomic spin-alignment component,  in good agreement with theory describing the interaction of collective spin operators and optical Stokes operators. 
Relative to earlier nonlinear strategies~\cite{Napolitano2011}, AOC allows increasing $\NL$ by an order of magnitude, giving 20 dB more signal and 10 dB less photon shot noise.
The resulting spin sensitivity surpasses by 9 dB the best-possible sensitivity of a linear $\mJy$ measurement with the same resources (photon number and allowed damage to the state).  
Theory shows that this advantage holds over all metrologically relevant conditions.

Understanding the limits of such nonlinear measurements has implications for instruments that naturally operate in nonlinear regimes, such as interferometers employing Bose-Einstein condensates~\cite{Gross2010,Luecke2011,Ockeloen2013} and optical magnetometers employing spin-exchange relaxation-free~\cite{Allred2002} and nonlinear~\cite{Budker2000a} magneto-optic rotation.
Similar nondestructive measurements are used in state-of-the-art optical magnetometers~\cite{Budker2007,Vasilakis2011a,Behbood2013b} and to detect the magnetization of spinor condensates~\cite{Sadler2006,Vengalattore2007,Liu2009a} and lattice gases~\cite{Brahms2011}, and are the basis for proposals for preparing~\cite{Toth2010a,Hauke2013,Behbood2013a,Puentes2013a} and detecting~\cite{Eckert2007,Eckert2008} exotic quantum phases of ultracold atoms.

\section{Nonlinear spin measurements}
\begin{figure}[t!]
	\centering
	\includegraphics[width=\columnwidth]{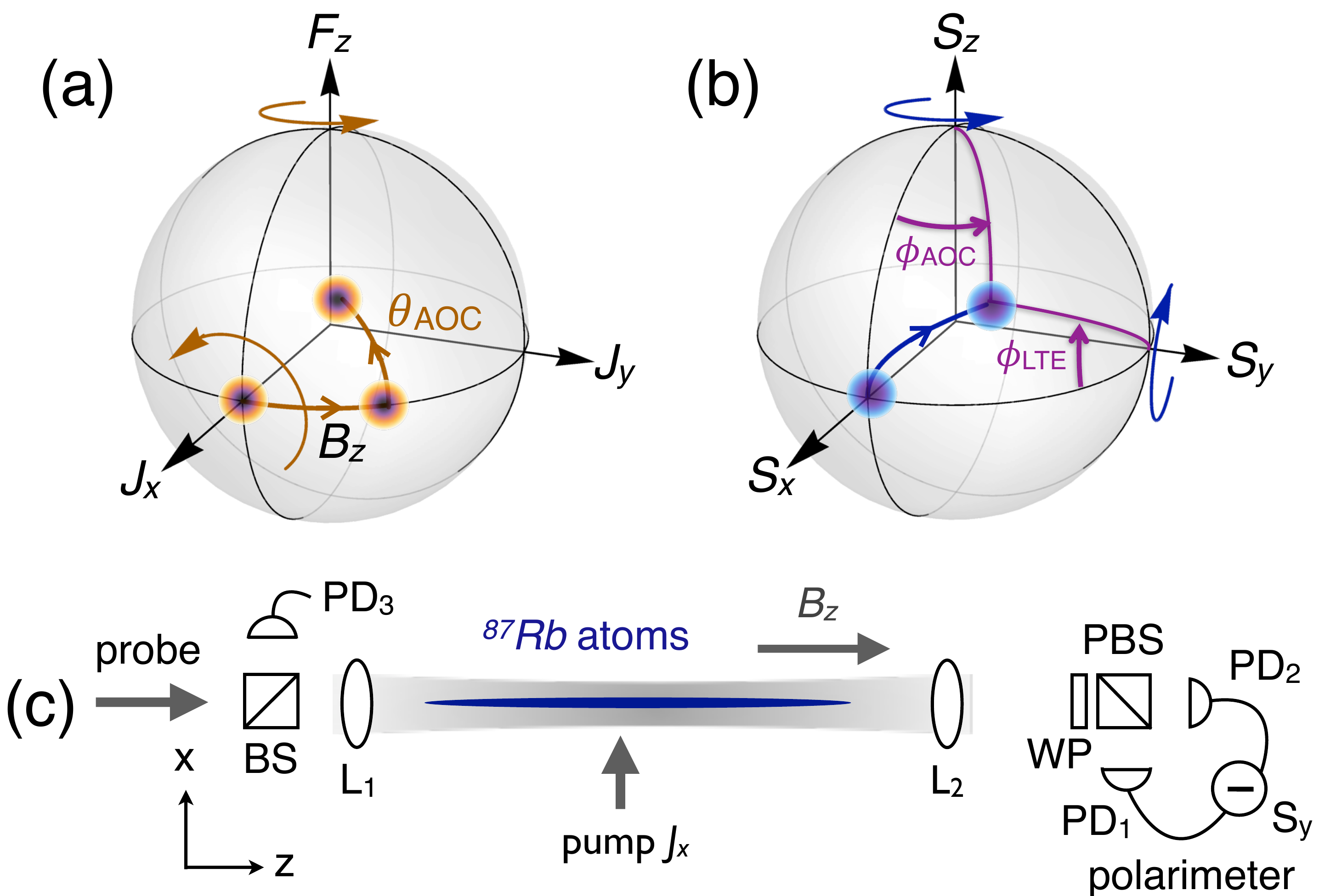}%
	\caption{
	Alignment-to-orientation conversion measurement of atomic spins.
	(a) An unknown field $B_z$ rotates an initially $\Jx$-polarized state in the $\Jx$-$\Jy$ plane.
	The $\Jy$ component is detected using an $\Sx$-polarized probe, which produces a rotation of $\Jy$ toward $\Fz$ by an angle ${\theta}_{\rm AOC} = \kB \Sx/2$.
	(b) Simultaneously, paramagnetic Faraday rotation produces a rotation of $\Sx$ toward $\Sy$.  
	The net effect is a rotation $\phiNL =  \kA \kB \NL \Jy/4$, which is observed by detecting $\Sy$.
	In an alternative strategy, the linear-to-elliptical rotation of $\Sx$ towards $\Sz$ by the angle $\phiL = \kB \Jy$ can be observed by detecting $\Sz$.
	(c) Experimental geometry.  
	Near-resonant probe pulses pass through a cold cloud of $^{87}$Rb atoms and experience a Faraday rotation by an angle proportional to the on-axis collective spin $\Fz$.
	Atoms are prepared in a coherent spin state $\Jx$ via optical pumping.  
	The pulses are initially polarized with maximal Stokes operator $\Sx$, measured at the input by photodiode (PD$_3$). 
	Rotation toward $\Sy$ is detected by a balanced polarimeter consisting of a waveplate (WP), polarizing beam splitter (PBS), and photodiodes (PD$_{1,2}$).
	\label{fig:experiment}
	}
\end{figure}

We work with an ensemble of  $\NA \sim 10^6$  laser-cooled $^{87}$Rb atoms held in an optical dipole trap, as illustrated in Fig.~\ref{fig:experiment}(a) and described in detail in Ref.~\cite{Kubasik2009}.
The atoms are prepared in the $f=1$ hyperfine ground state, and interact dispersively with light pulses of duration $\tau$ via an effective Hamiltonian~\cite{Echaniz2005}
\begin{equation}
	 \HH = \kA\Fz\tSz + \kB(\Jx\tSx + \Jy\tSy) -  \gamma_F {\bf B} \cdot {\bf F},
	\label{eqn:H_full}
\end{equation}
where the coupling coefficients $\kAB$ are proportional to the vectorial and tensorial polarizability, respectively, and $\gamma_F$ is the {ground-state} gyromagnetic ratio. 
Here, the operators $\Ji$ describe the collective atomic spin, and the optical polarization is described by the pulse-integrated Stokes operators $ \int dt \, \tS_i(t) \equiv \Si$ (see Appendix~A).
$\Jx$ and $\Jy$ represent the collective spin alignment, i.e., Raman coherences between states with $\Delta m_f=2$, and $\Fz$ describes the collective spin orientation along the quantization axis, set by the direction of propagation of the probe pulses.
$\Sx$ and $\Sy$ describe linear polarizations, while $\Sz$ is the degree of circular polarization, i.e., the ellipticity.

In regular Faraday rotation, the collective spin orientation $\Fz$ is detected indirectly by measuring the polarization rotation of an input optical pulse due to the first term in Eq.~(\ref{eqn:H_full}).
Typically, the input optical pulse is $\Sx$-polarized, and the polarization rotation is detected in the $\Sy$ basis.
Detection of the collective spin alignment $\mJy$ (or $\mJx$) requires making use of the second term in Eq.~(\ref{eqn:H_full}), and can be achieved with either a linear or a nonlinear measurement strategy, as we now describe (see Figure~\ref{fig:experiment}).

In a linear measurement the polarization rotation due to the term $\kB \Sy\Jy$ is directly measured -- e.g., an input $\Sx$-polarized probe (i.e. $\ave{\Sx}=\NL/2$) is rotated toward $\Sz$ by a small angle $\phiL = \kB \Jy$.
We refer to this type of strategy as linear-to-elliptical (LTE) measurement of $\mJy$.
It gives quantum-limited sensitivity 
\begin{equation}
\label{eqn:LTEScaling}
(\Delta \mJy)^2_{\rm \LIN} = \frac{(\Delta\Sz\supin)^2}{\kB^2\mSx^2} = \frac{1}{\kB^2} \frac{1}{\NL},
\end{equation}
i.e., with shot-noise scaling.
Here and in the following we use the notation $\mJy\equiv\ave{\Jy}$ for expectation values.
The same sensitivity is achieved with other linear measurement strategies employing different input polarizations.
Note that for large detunings, $\kA\propto\Delta^{-1} \gg \kB\propto\Delta^{-2}$, so detection of $\mJy$ with this method is less sensitive than regular Faraday-rotation detection of $\Fz$.

AOC measurement of $\mJy$ employs $\HH$ twice and gives a signal nonlinear in $\NL$: 
The term $\kB\Sx\Jx$ produces a rotation of $\Jy$ toward $\Fz$ by an angle ${\theta}_{\rm AOC} = \kB \mSx/2$.  
Simultaneously, the term $\kA\Sz\Fz$ produces a rotation of $\Sx$ toward $\Sy$ by an angle $\phiNL = \kA \Fz$.  
The net effect is an optical rotation $\phiNL =  \kA \kB \NL \Jy/4$, which is observed by detecting $\Sy$.  
The quantum-noise-limited sensitivity of this nonlinear measurement is (see Appendix~B)
\begin{equation}
\label{eqn:AOCScaling}
(\Delta \mJy)^2_{\rm AOC} =  (\frac{4}{\kB\NL})^2(\frac{1}{\kA^2\NL}+\frac{\NA}{4})
\end{equation}
with scaling $\Delta \mJy \propto \NL^{-3/2}$ crossing over to $\Delta\mJy\propto\NL^{-1}$ at large $\NL$.
Using the Hamiltonian in second order, AOC gives a signal $\propto \kA \kB \NL$, versus $\propto  \kB$ for LTE, which is an advantage at large detuning, where $\kA \gg \kB$.

Both strategies employ the same measurement resources, namely, an $\Sx$-polarized coherent-state probe, so that the quantum uncertainties on the input-polarization angles are $\Delta \Sy\supin/\mSx = \Delta \Sz\supin/\mSx = \NL^{-1/2}$.
In addition to the coherent  rotations produced by $\HH$, spontaneous scattering of probe photons causes two kinds of ``damage'' to the spin state: loss of polarization (decoherence) and added spin noise.  
The tradeoff between information gain and damage is what ultimately limits the sensitivity of the $\mJy$ measurement~\cite{Echaniz2005,WolfgrammNPhot2013,SewellNatPhot2013}.  
For equal $\NL$, the damage is the same for the {\LIN} and AOC measurements, because they have the same initial conditions and differ only in whether $\Sz$ or $\Sy$ is detected.

\begin{figure}[t!]
	\centering
	\includegraphics[width=\columnwidth]{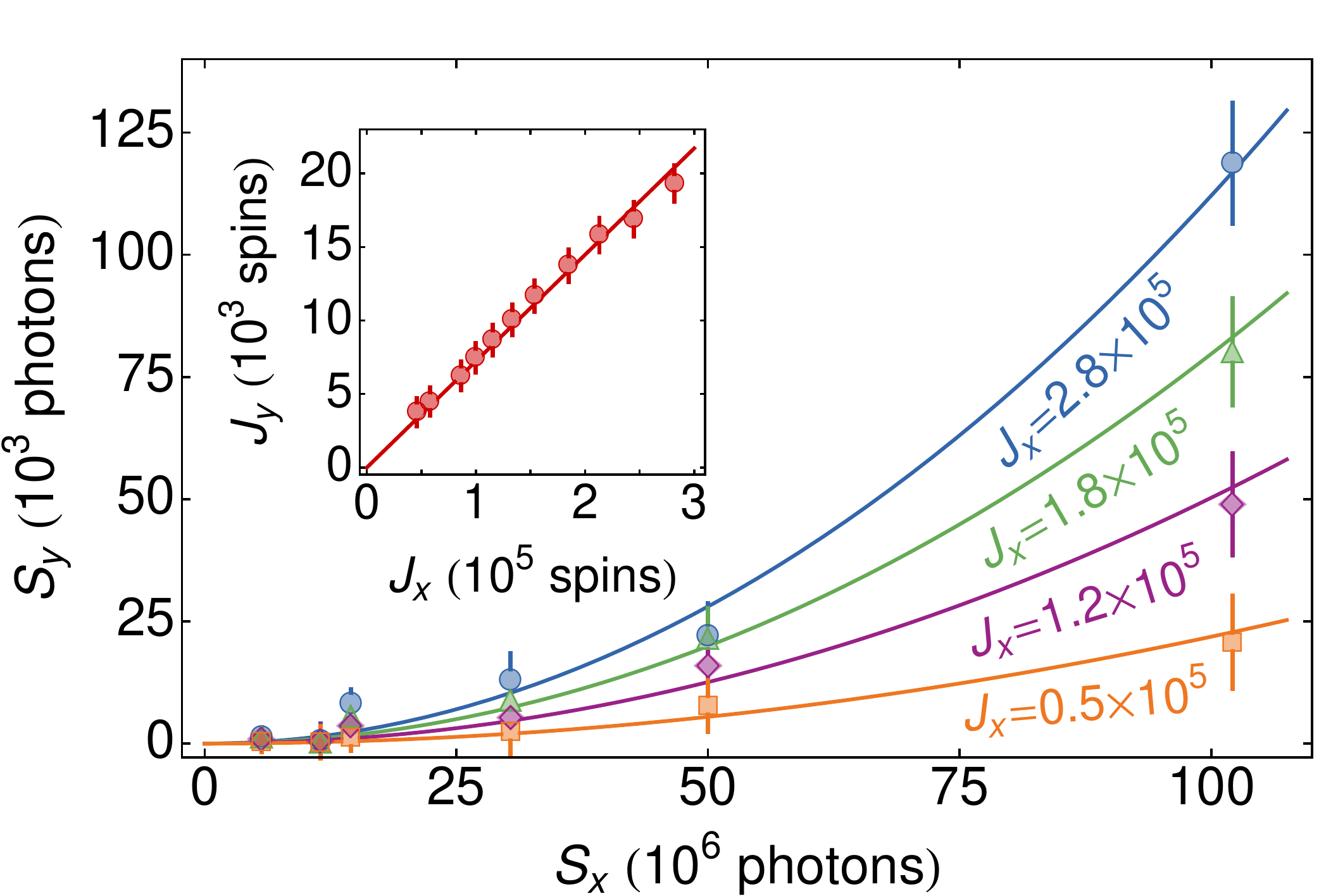}%
	\caption{
	Alignment-to-orientation conversion measurement of $\mJy$.
	In the main frame we plot the signal $\mSy=\ave{\Sy\supout}$ of the AOC measurement as a function of $\mSx$ for various $\mJx$.
	We find the measured signal $\mJy$ from fit to data using the function $\mSy=(\kA\kB/2) \mJy \mSx^2$ (solid lines).
	Error bars represent $\pm1\sigma$ statistical errors.
	Inset: Measured $\mJy$ versus $\mJx$. 
	For small rotation angles $\mJy\simeq2\omega_L B_z t \mJx$.
	\label{fig:signal}
	}
\end{figure}

From these scaling considerations, the AOC measurement should surpass the {\LIN} measurement in sensitivity, 
$(\Delta \mJy)^2_{\rm AOC} <(\Delta \mJy)^2_{\rm \LIN}$ for $\NL \gtrsim 16/(\kA^2 \NL)+ 4\NA$, but only if such a large $\NL$ 
does not cause excessive scattering damage to $\mJy$. In atomic ensembles, the achievable information-damage 
tradeoff is determined by the optical depth (OD)~\cite{Echaniz2005}, which, in principle, can grow without bound.  
For high-OD ensembles the nonlinear measurement will, through advantageous scaling, surpass the best-possible linear measurement of the same quantity, under the same conditions.  
In what follows, we confirm this prediction experimentally, by comparing measured AOC sensitivity to the calculated best-possible sensitivity of the {\LIN} measurement.

\begin{figure}[t!]
	\centering
	\includegraphics[width=\columnwidth]{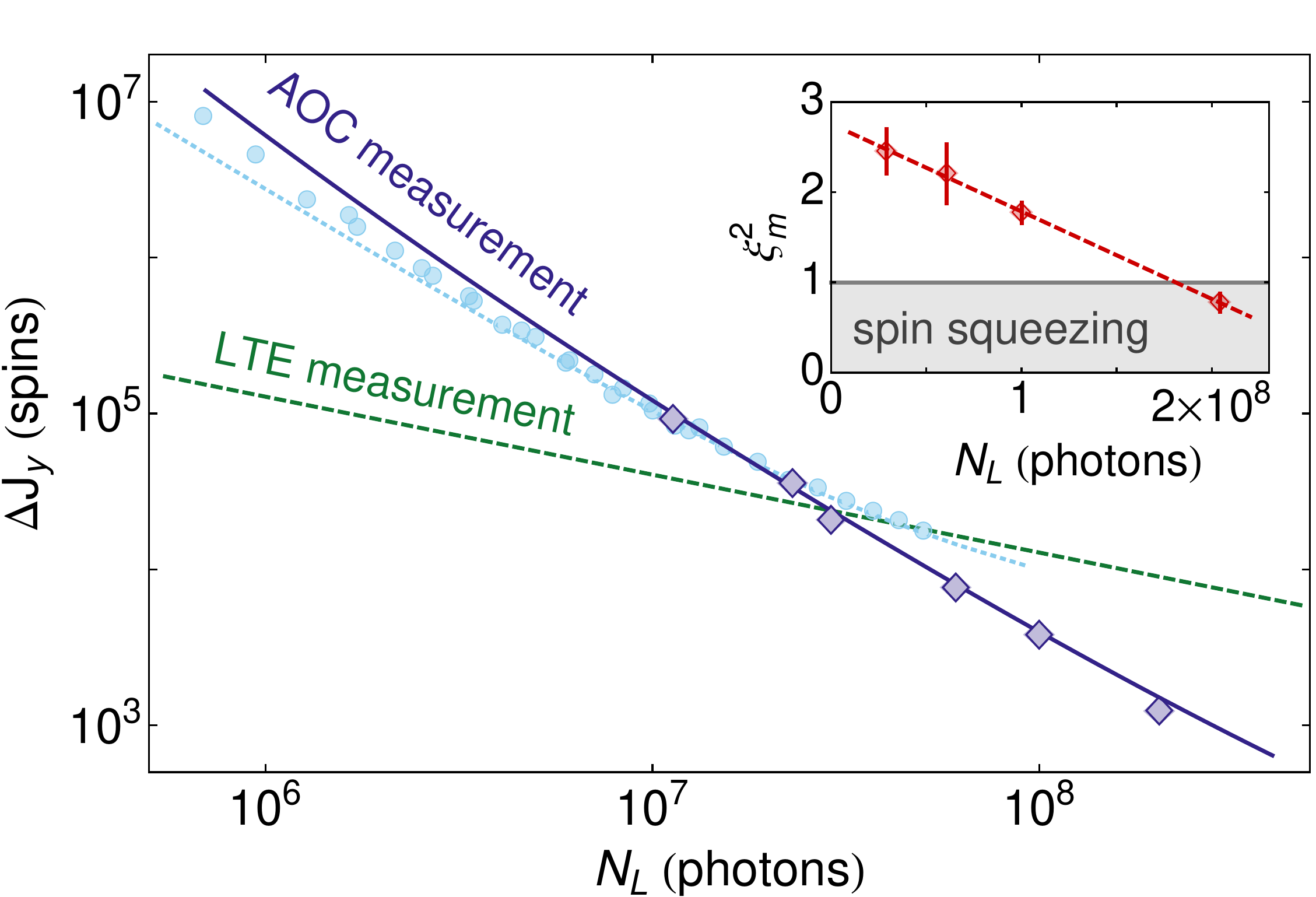}%
	\caption{
	Log-log plot of the uncertainty $\Delta\mJy$ of the AOC measurement versus number of photons $\NL$.
	Blue diamonds indicate the measured sensitivity.
	Nonlinear enhanced scaling of the sensitivity is observed over more than an order of magnitude in $\NL$.
	A fit to the data yields $\Delta\mJy\propto\NL^k$ with $k=-1.46\pm0.04$.
	The best observed sensitivity is $\Delta\mJy=1290\pm90$ spins with $\NL=2\times10^8$ photons.
	For reference, we also plot the data (light blue circles) and theory (dotted curve) for the measurement of $\Fz$ via nonlinear Faraday rotation reported in Napolitano et al.~\cite{Napolitano2011}
	The solid blue curve shows the theoretical prediction given by Eq.(\ref{eqn:AOCScaling}) with no free parameters, plus the independently measured electronic noise contribution.
	Dashed green curve: theory curve describing an ideal LTE measurement of $\mJy$ without technical or electronic noise contributions.
	The nonlinear measurement sensitivity surpasses an ideal LTE measurement with $\NL=3\times10^7$ photons.
	Error bars for standard errors would be smaller than the symbols and are not shown.
	Inset: Observed metrologically significant spin squeezing $\xi_m^2$ as a function of photon number.
	The dashed line is a guide to the eye.
	Error bars indicate $\pm1\sigma$ standard errors.
	\label{fig:scaling}
	}
\end{figure}

\section{Experimental data and analysis}
The experimental system, illustrated in Fig.~\ref{fig:experiment}(c), is the same as in Ref.~\cite{Sewell2012}, with full details given in Ref.~\cite{Kubasik2009}.
After loading up to $6\times10^5$ laser-cooled atoms into a single-beam optical dipole trap, we prepare a $\Jx$-aligned coherent spin state (CSS) via optical pumping, $\mJx=\ave{\Jx}=\NA/2$.  
An (unknown) bias field $\Bz$ rotates the state in the $\Jx$-$\Jy$  plane at a rate $2\,\omega_L$, where $\omega_L=-\gamma_F B_z$ is the Larmor frequency, to produce $\mJy = \ave{\Jy} =\sin(2\,\omega_L  t) \mJx$, which we then detect via AOC measurement.
We probe the atoms with a sequence of $2$-$\mu$s-long-pulses of light sent through the atoms at $5$-$\mu$s intervals and record $\Sy\supout$ with a shot-noise-limited balanced polarimeter.
The pulses have a detuning $\Delta/2\pi=-600$ MHz, i.e. to the red of the $F=1\rightarrow F'=0$ transition on the $\dd$ line.
To study noise scaling we vary both the number of photons per pulse $\NL$, and the number of atoms $\NA$ in the initial coherent spin state.

In Fig.~\ref{fig:signal} we plot the observed signal $\mSy=\ave{\Sy\supout}$ versus $\mSx$ for various values of $\mJx$.
As expected, we observe a signal that increases quadratically with $\mSx$.
We extract $\mJy$ from a fit to data using the function $\mSy=(\kA\kB/2) \mJy \mSx^2$ (solid lines in Fig.~\ref{fig:signal}), where the coupling constants $\kA=1.47\times10^{-7}$~rad/spin and $\kB=7.54\times10^{-9}$~rad/spin are independently measured~\cite{Sewell2012}
In the inset we plot the measured $\mJy$ versus $\mJx$.
For small rotation angles $\mJy\simeq2\,\omega_L t \, \mJx$, where $t=7.5$~$\mu$s is the time between the centers of the baseline and AOC measurements.
A linear fit to the data yields $B_z=\unit{103\pm3}{\nano\tesla}$.

The measured sensitivity $\Delta\mJy=\Delta\Sy /( \kA\kB \mSx^2) $ is obtained from the measured readout variation $\Delta\Sy$ and the slope $\partial \mSy/\partial \mJy = \kA\kB \mSx^2$, with the contribution due to the atomic projection noise subtracted (see Appendix B).
As shown in Fig.~\ref{fig:scaling}, we observe nonlinear enhanced scaling $\Delta\mJy\propto\NL^{-3/2}$ over more than an order of magnitude in $\NL$.
For these data, $\mJx=2.8\times10^5$ and $\mJy=1.9\times10^4$.
The data are well described by the theoretical model of Eq.~(\ref{eqn:AOCScaling}), plus a small offset due to electronic noise, which is independently measured (see Appendix~C).
We observe a minimum $\Delta\mJy=1230\pm90$ spins with $\NL=2\times10^8$ photons.

The AOC measurement sensitivity crosses below the ideal LTE measurement $(\Delta \mJy)_{\rm \LIN} = (1/\kB) \NL^{-1/2}$ (dashed green line in Fig.~\ref{fig:scaling}) with $\NL=3\times10^7$ photons, indicating that, for our experimental parameters, the nonlinear measurement is the superior measurement of $\Jy$.
For comparison, we also compare our measurement of the alignment $\mJy$ with the nonlinear Faraday-rotation measurement of $\mFz$ reported in Napolitano et al.~\cite{Napolitano2011} (light blue circles and dotted line in Fig.~\ref{fig:scaling}).
We note, in particular, that the advantageous scaling of the current measurement extends to an order-of-magnitude larger $\NL$ than reported in that work.

Nondestructive, projection-noise-limited measurement can be used to prepare a conditional spin-squeezed atomic state~\cite{Kuzmich1998}.
Generation of squeezing is a useful metric for the measurement sensitivity since it takes into account damage done to the atomic state by the optical probe~\cite{Wineland1992,SewellNatPhot2013}.
Here, it is important to note that although the AOC signal is proportional to the atomic spin alignment $\mJy$, quantum noise from the spin orientation $\Fz$ is mixed into the measurement: Scaled to have units of spins, the Faraday-rotation signal from the AOC measurement is $\hat{\Phi}\equiv(\cos\theta/\kA\mSx)\Sy\supout=(\cos\theta/\kA\mSx)\Sy\supin + \K\supin$, which describes a nondestructive measurement of the mixed alignment-orientation variable $\K\supin\equiv\Fz\supin\cos\theta+\Jy\supin\sin\theta$, where $\tan\theta\equiv\kB\mSx/2$ (see Appendix A).
$\K$ is the variable that should be squeezed to enhance the sensitivity of the AOC measurement.
Metrological enhancement is quantified by the spin-squeezing parameter $\xi_m^2 \equiv (\Delta\K\supout)^2\mJx/2\norm{\mJx\supout}^2$~\cite{Wineland1992} (see Appendix D).
With $\NL=2\times10^8$ and $\mJx=2.8\times 10^{5}$, we observe a conditional noise $2.3\pm0.5$ dB below the projection noise limit and $\xi_m^2 =0.7\pm0.2$, or $1.5\pm0.8$ dB of metrologically significant spin squeezing (inset of Fig.~\ref{fig:scaling}).
We note that for our experimental parameters, LTE would not induce spin squeezing.

\section{Discussion}
The experiment shows AOC surpassing LTE through improved scaling at the specific detuning of $\Delta/2\pi=-600$ MHz.  
It is important to ask whether this advantage persists under other measurement conditions.
A good metric for the optimum measurement is the number of photons $\NL$ required to achieve a given sensitivity (see Appendix~E).
In Fig.~\ref{fig:sen}(a) we plot the calculated $\NL$ required to reach projection-noise-limited sensitivity for the two measurement strategies, i.e. $(\Delta \mJy)^2_{\rm AOC} = (\Delta \mJy)^2_{\rm \LIN} = \NA/4$ for our experimental parameters.
For comparison, we also plot curves showing the damage $\eta_{\rm sc}$ to the atomic state due to spontaneous emission.
We see that the AOC strategy achieves the same sensitivity with fewer probe photons (and thus causes less damage) except very close to the atomic resonances, i.e. except in regions where large scattering rates make the quantum nondemolition measurement impossible anyway.
Another important metric is the achievable metrologically significant squeezing, found by optimizing $\xi_m^2$ over  $\NL$ at any given detuning.
In Fig.~\ref{fig:sen}(b) we show this optimal $\xi_m^2$ versus detuning.
The global optimum squeezing achieved by the AOC (LTE) strategy is $\xi_m^2=0.47$ (0.63) at a detuning of $\Delta/2\pi=-59$~MHz (77~MHz).

\begin{figure}[t!]
	\centering
	\includegraphics[width=\columnwidth]{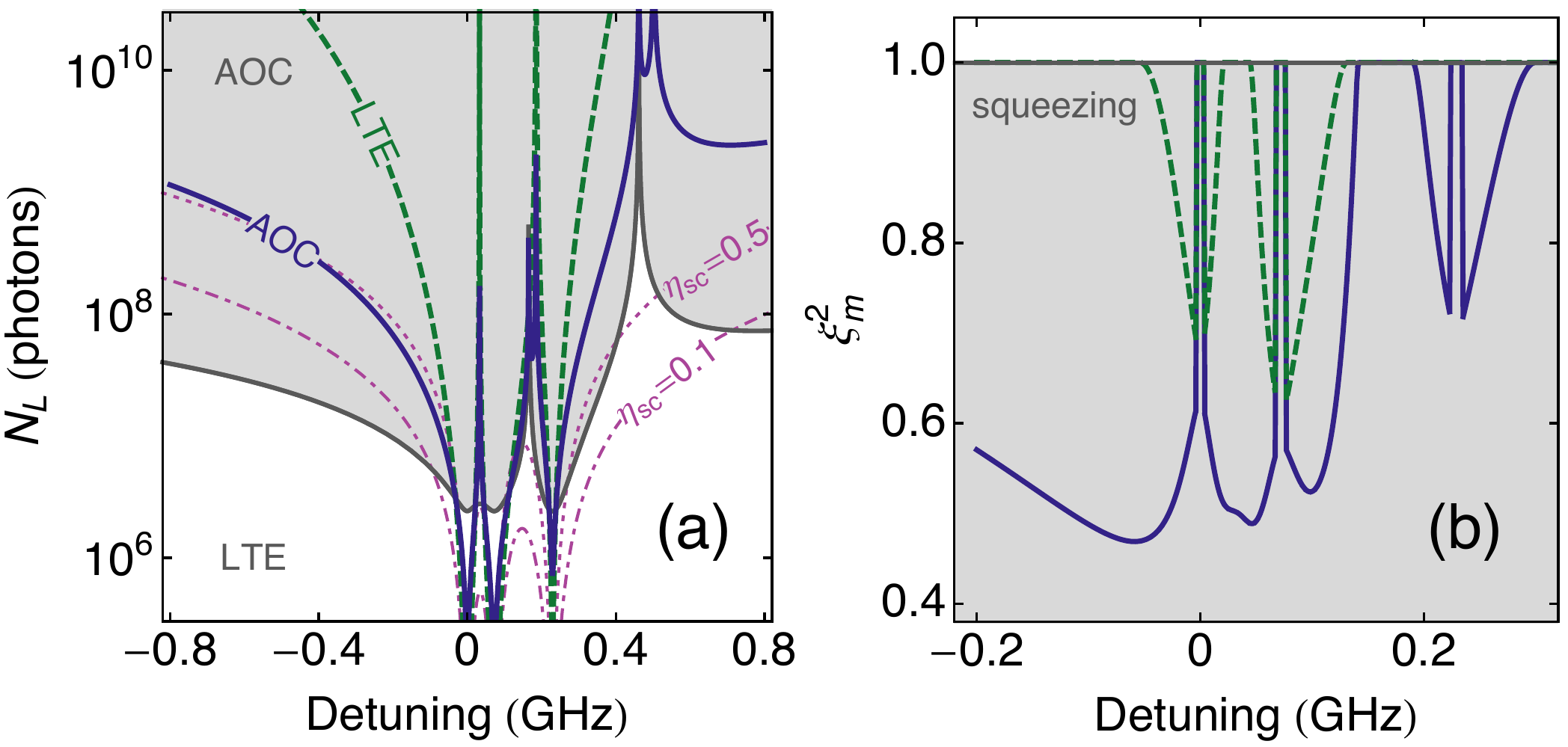}%
	\caption{ 
	Theoretical comparison of AOC (solid blue curves) and LTE (dashed green curves) measurement sensitivity.
	(a) Number of photons $\NL$ needed to achieve projection-noise-limited sensitivity $(\Delta \Jy)^2_{\rm AOC} = (\Delta \Jy)^2_{\rm \LIN} = \NA/4$ as a function of detuning $\Delta$.
	The gray line indicates $(\Delta \Jy)^2_{\rm AOC}=(\Delta \Jy)^2_{\rm \LIN}$, so that the AOC (LTE) strategy is more sensitive in the shaded (white) region.
	Magenta curves represent damage $\eta_{\rm sc}=0.1$ (dot-dashed) and 0.5 (dotted) to the atomic state due to spontaneous emission. 
	(b) Estimated metrologically significant spin squeezing $\xi_m^2$, optimized as a function of $\NL$, versus probe detuning.
	\label{fig:sen}
	}
\end{figure}

\begin{figure*}[t!]
	\centering
	\includegraphics[width=\textwidth]{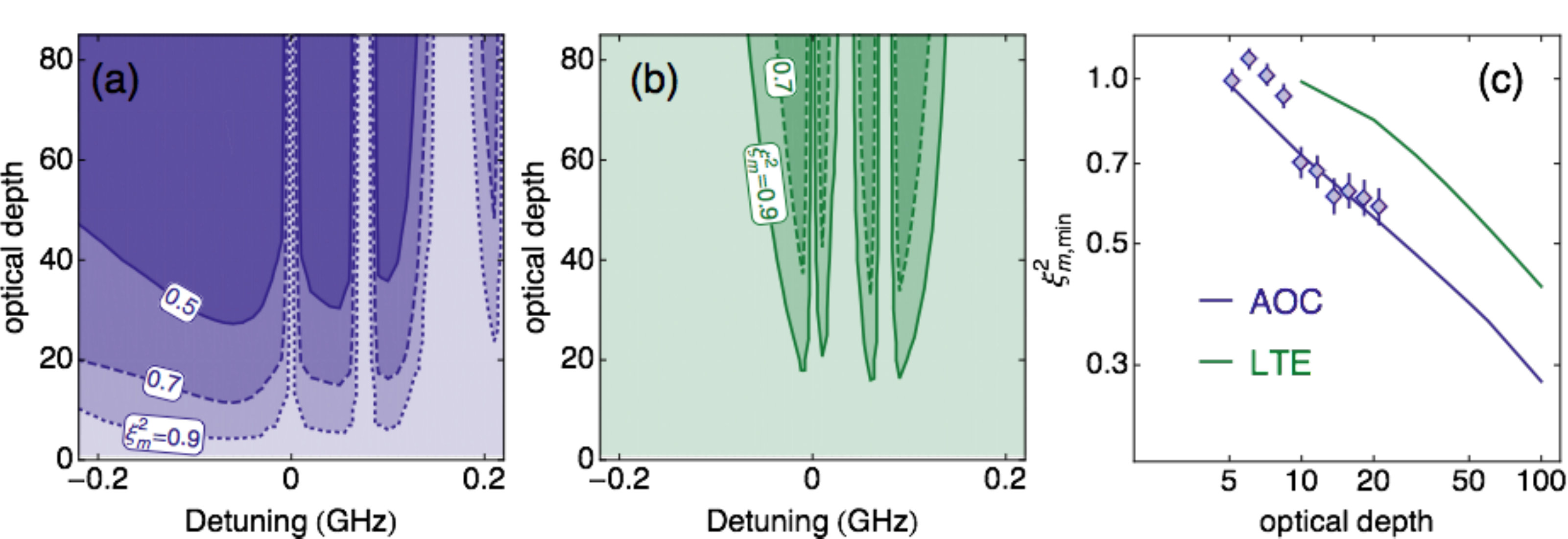}%
	\caption{
	Theoretical calculation of spin squeezing as a function of optical depth and probe detuning for (a) the AOC strategy and (b) the LTE strategy with our experimental parameters.
	Contours indicate the minimum achievable metrologically significant squeezing $\xi_{m,min}^2$ with respect to $\NL$, with values indicated, as a function of detuning $\Delta$ and optical depth $\OD$.
	(c) The blue diamonds represent the observed spin squeezing $\xi_m$ as a function of optical depth for the AOC measurement with $\NL=2\times10^8$~photons.
	The solid curves show the theoretically predicted minimum achievable squeezing, optimized with respect to both $\NL$ and $\Delta$, for the AOC measurement (solid blue line) and the LTE measurement (solid green line).
	The scaling of each curve is roughly $\xi^2 \propto {\rm OD}^{-1/2}$, so the advantage for AOC continues also to large OD. 
 	\label{fig:sq}
	}
\end{figure*}

In Fig.~\ref{fig:sq}, we plot the achievable $\xi_{m,min}^2$ as a function of $\NL$ versus both detuning $\Delta$ and optical depth $\OD$ for the AOC [Fig.~\ref{fig:sq}(a)] and LTE  [Fig.~\ref{fig:sq}(b)] strategies.
We find that AOC is globally optimum, giving more squeezing, and thus better metrological sensitivity, across the entire parameter range.
In Fig.~\ref{fig:sq}(c), we plot the fully optimized spin squeezing, i.e., over $\Delta$ and $\NL$, achievable by the AOC and LTE measurement strategies as a function of OD.   
This comparison again shows an advantage for AOC, including for large OD, and agrees well with experimental results.

We conclude that:
(1) for nearly all probe detunings, if  $\NL$  is chosen to give projection-noise sensitivity for LTE, then AOC gives better sensitivity at the same detuning and $N_L$.  
The exception is probing very near an absorption resonance, which induces a large decoherence in the atomic state.  
(2) Considering as a figure of merit the achievable spin squeezing, or equivalently, the magnetometric sensitivity of a Ramsey sequence employing these measurements~\cite{Sewell2012}, the global optimum, including choice of measurement, is AOC at a detuning of $-59$MHz, with $\NL=5.4\times10^6$ photons.  
In this practical metrological sense, the nonlinear measurement is unambiguously superior.
Although the AOC and LTE compared here use coherent states as inputs, the same conclusion is expected when nonclassical probe states are used:  
For both measurements, the optical rotation sensitivity $\Delta S_{\rm y,z}\supin/\mSx$  can be enhanced in the same way by squeezing~\cite{Wolfgramm2010} and other techniques~\cite{WolfgrammNPhot2013}.

\section{Conclusion}
We have identified a scenario -- nondestructive detection of atomic spin alignment -- in which a nonlinear measurement (AOC) outperforms competing linear strategies with the same experimental resources.
Our experimental demonstration answers a fundamental question in quantum metrology~\cite{JavanainenPRA2012,Zwierz2010,Demkowicz-Dobrzanski2012,Hall2012a}, with implications for quantum enhancement of atomic instruments operating in nonlinear regimes~\cite{Budker2000a,Allred2002,Gross2010}.
Beyond magnetometry, our techniques may be useful in the measurement of spinor condensates~\cite{Sadler2006,Vengalattore2007,Liu2009a} and lattice gases~\cite{Brahms2011}. 
To date, such measurements have been limited to detecting spin orientation (vector magnetization), whereas our technique provides a nondestructive measurement of spin alignment (a component of the spin-one nematic tensor), with direct application, e.g., to the detection of spin-nematic quadrature squeezing in spinor condensates~\cite{Hamley2012}.
The technique may make possible proposals for the detection~\cite{Eckert2007,Eckert2008} and preparation~\cite{Hauke2013,Behbood2013a} of exotic quantum phases of ultracold atoms, which require quantum-noise-limited measurement sensitivity.


\appendix

\section{Atom-light interaction}
As described in Refs.~\cite{Echaniz2005,Colangelo2013a}, the light pulses and atoms interact by the effective Hamiltonian
\begin{equation}
	 \HH =  \kA\Fz\tSz(t) + \kB[\Jx\tSx(t) + \Jy\tSy(t)],
	\label{supp:H_full}
\end{equation}
plus higher-order terms describing fast electronic nonlinearities~\cite{NapolitanoNJP2010}.
Here, $\kAB$ are coupling constants that depend on the beam geometry, excited-state linewidth, laser detuning, and the hyperfine structure of the atom, and the light is described by the time-resolved Stokes operator $\btS(t)$, defined as   $ \tS_i \equiv \frac{1}{2} ({\cal E}\supminus_+ ,{\cal E}\supminus_-) \sigma_i ({\cal E}\supplus_+ ,{\cal E}\supplus_-)^T $, where the $ \sigma_i $ are the Pauli matrices and ${\cal E}\supplus_\pm(t)$ are the positive-frequency parts of  quantized fields for the circular plus or minus polarizations.  
The pulse-averaged Stokes operators are $\Si \equiv \int dt\, \tS_i(t)$ so that $\Si = \frac{1}{2} (a^\dagger_+ ,a^\dagger_-) \sigma_i (a_+ ,a_-)^T $, where  $a_\pm$ are operators for the temporal mode of the pulse~\cite{Sewell2012}.
In all scenarios of interest $\ave{\Jx} \approx \NA/2 \gg \ave{\Jy},\ave{\Fz}$, and we use input $\Sx$-polarized light pulses, $\mSx=\ave{\Sx\supin}=\NL/2$, and detect the output $\Sy\supout$ component of the optical polarization.

The atomic spin ensemble is characterized by the operators $\Fz\equiv\sum_{i}^{\NA}\fz^{(i)}/2$, describing the collective spin orientation, and $\Jxy\equiv\sum_{i}^{\NA}\jxy$, describing the collective spin alignment, where $\jx \equiv (\myf_x^2 - \myf_y^2)/2 $ and $\jy \equiv (\myf_x\myf_y + \myf_y\myf_x)/2 $ describe single-atom Raman coherences, i.e., coherences between states with $\Delta m_f=2$.
Here, ${\bf f}^{(i)}$ is the total spin of the $i$~th atom.
For $f=1$, these operators obey commutation relations $[\Jx,\Jy]= i\Fz$ and cyclic permutations.

Using Eq.~(\ref{supp:H_full})  in the Heisenberg equations of motion and integrating over the duration of a single light pulse~\cite{Sewell2012}, we find the detected outputs to second order in $\Sx$
\begin{align}
\label{eqn:LTESignal}
          \Sz\supout &= \Sz\supin+\kB\Sy\Jx\supin-\kB\Sx\Jy\supin \\
\label{eqn:AOCSignal}
	 \Sy\supout &= \Sy\supin+\kA\Sx \Fz\supin+\frac{\kA \kB}{2}\Sx^2\Jy\supin \nonumber \\
	 	&= \Sy\supin + \frac{\kA\Sx}{\cos\theta} \K\supin
\end{align}
plus small terms. 
Equation~(\ref{eqn:AOCSignal}) describes a nondestructive measurement of the mixed alignment-orientation variable $\K\supin\equiv\Fz\supin\cos\theta+\Jy\supin\sin\theta$, where $\tan\theta\equiv\kB\mSx/2$.
$\K$ is the variable that should be squeezed to enhance the sensitivity of the AOC measurement.

\section{Measurement sensitivity}
Both AOC and LTE measurements have the same input state, with $\mSx\equiv\ave{\Sx\supin}=\NL/2$, $\mJx = \ave{\Jx} = \NA/2$, $\ave{\Fz}=0$, $(\Delta\Sy\supin)^2=\NL/4$, $(\Delta\Fz\supin)^2=(\Delta\Jy\supin)^2=\NA/4$, and uncorrelated $\Sx, \Sy, \Sz, \Fz,$ and $\Jy$.  

The LTE measurement detects $\Sz$, with signal $\ave{\Sz\supout}=-\kB\ave{\Sx}\ave{\Jy}$ and variance $(\Delta\Sz\supout)^2= (\Delta\Sz\supin)^2 + \kB^2\ave{\Jx}^2 (\Sy\supin)^2 + \kB^2\ave{\Sx}^2 (\Delta\Jy\supin)^2$. 
To infer the $\Jy$ measurement uncertainty, we note that $\Jy$ and $\Sz\supout$ are Gaussian variables \cite{MadsenPRA2004}, so that the simple error-propagation formula coincides with more sophisticated estimation methods using, e.g., Fisher information \cite{Pezze2006a}.  
We find:
\begin{align}
	(\Delta\ave\Jy)^2&=\frac{(\Delta\Sz\supout)^2}{|\partial\ave{\Sz\supout}/\partial\ave{\Jy}|^2} \\
	&=\frac{1}{\kB^2\NL}+\frac{\NA^2}{4  \NL}+\frac{\NA}{4},
\end{align}
which shows shot-noise scaling.   
The first two terms are readout noise and determine the measurement sensitivity. 
In the experiment $(\kB \NA)^2/4 \sim 10^{-3}$, so the second term is negligible. 
The last term is due to the variance of  $\Jy$ -- i.e., the signal we are trying to estimate -- which we subtract to give the expression in Eq.~(\ref{eqn:LTEScaling}).
We note that other measurement strategies using the same term in the Hamiltonian are possible, e.g., probing with $\Sz$-polarized light and reading out the rotation of $\Sz$ into $\Sy$, but lead to the same measurement sensitivity.

The AOC measurement detects $\Sy$, with signal $\ave{\Sy\supout}=(\kA\kB\ave{\Sx}^2/2)\ave{\Jy}$ and variance $(\Delta\Sy\supout)^2= (\Delta\Sy\supin)^2 + (\kA^2\ave{\Sx}^2)(\Delta\Fz\supin)^2 + (\kA^2\kB^2\ave{\Sx}^4/4)(\Delta\Jy\supin)^2$. 
From this slope and output variance we find the variance referred to the input
\begin{align}
	(\Delta\ave\Jy)^2 &= \frac{(\Delta \Sy\supout)^2}{|\partial\ave{\Sy\supout}/\partial\ave{\Jy}|^2} \\
\label{eqn:JyNoise}
	& = \frac{16}{\kA^2\kB^2\NL^3} + \frac{4 \NA}{\kB^2\NL^2} + \frac{\NA}{4},
\end{align}
where again the last term is the signal variance, which we subtract to give the expression in Eq.~(\ref{eqn:AOCScaling}).

In contrast, previous work~\cite{Napolitano2011} used short, intense pulses to access a nonlinear term in the effective Hamiltonian $\kappa_{\rm NL} S_0 \Sz \Fz$.  
The coupling $\kappa_{\rm NL}$ is proportional to the Kerr nonlinear polarizability, and $S_0\equiv\NL/2$ (so that $S_0 = \Sx$ for the input polarization used).  
Calculating the variance referred to the input as above, we find sensitivity $\Delta \mFz = \Delta\Sz\supin/(\kappa_{\rm NL} \mSx^2)$ and $\Delta \mFz \propto \NL^{-3/2}$ scaling.  

\section{Electronic \& technical noise}
The measured electronic noise of the detector referred to the interferometer input is $\EN=9.2\times10^5$ photons, and contributes a term $\EN \times 64/(\kA^2\kB^2\NL^4)$ to Eq.(\ref{eqn:AOCScaling}), which is included in the blue curve plotted in Fig.~\ref{fig:scaling}.
Technical noise contributions from both the atomic and light variables are negligible in this experiment.  
Baseline subtraction is used to remove low-frequency noise.

\section{Conditional noise reduction \& spin squeezing}
Measurement-induced noise reduction is  quantified by the conditional variance $\var(\K|\Phi_1)=\var(\Phi_2-\chi\Phi_1)-\var(\phN)$, 
$\hat{\Phi}\equiv(\cos\theta/\kA\mSx)\Sy\supout=(\cos\theta/\kA\mSx)\Sy\supin + \K$, and $\chi\equiv\cov(\Phi_1,\Phi_2)/\var(\Phi_1) > 0$.
Spin squeezing is quantified by the Wineland criterion~\cite{Wineland1992}, which accounts for both the noise and the coherence of the postmeasurement state: if $\xi_m^2 \equiv 2(\Delta\K)^2 \mJx /(\mJx\supout)^2$, where $\mJx\supout=(1-\eta_{\rm sc})(1-\eta_{\rm dep})\mJx$ is the mean alignment of the state after the measurement, then $\xi_m^2<1$ indicates a metrological advantage.
For this experiment, the independently measured depolarization due to probe scattering and field inhomogeneities give $\eta_{\rm sc}=0.093$ and $\eta_{\rm dep}=0.034$, respectively~\cite{Sewell2012}.
The subtracted noise contribution with $\NL=2\times10^8$photons is $\var(\phN)=1.3\times10^5$~spins$^2$.

\section{Dependence on detuning and optical depth}
The detuning dependence of the coupling constants $\kA$ and $\kB$ of Eq.(\ref{eqn:H_full}) is given by
\begin{align}
	\kA &= \frac{\sigma_0}{A} \frac{\Gamma}{16}\left[-4\delta_0(\Delta)-5\delta_1(\Delta)+5\delta_2(\Delta)\right] \\
	\kB &= \frac{\sigma_0}{A} \frac{\Gamma}{16}\left[4\delta_0(\Delta)-5\delta_1(\Delta)+\delta_2(\Delta)\right]
\end{align}
where $\delta_i(\Delta)\equiv1/\sqrt{\Gamma^2+(\Delta-\Delta_i)^2}$, $\Delta_i$ is the detuning from resonance with the $F=1\rightarrow F'=i$ transition on the $^{87}$Rb $D_2$ line, $\Gamma/2\pi=6.1$~MHz is the natural linewidth of the transition, $\Delta$ is measured from the $F=1\rightarrow F'=0$ transition, $\sigma_0\equiv\lambda^2/\pi$ and $A=4.1\times10^{-9}\,{\rm m}^2$ is the effective atom-light interaction area.
Note that for large detuning, i.e., $\Delta\gg\Gamma$, $\kA\propto1/\Delta$ and $\kB\propto1/\Delta^2$.

At any detuning, the measurement sensitivity can be improved by increasing the number of photons $\NL$ used in the measurement.
Note, however, that increasing $\NL$ also increases the damage $\eta_{\rm sc}=k(\Delta)\eta_{\gamma}(\Delta) \NL$ done to the atomic state we are trying to measure due to probe scattering, where $\eta_{\gamma}(\Delta)$ is the probability of scattering a single photon:
\begin{align}
	\eta_{\gamma} &= \frac{\sigma_0}{A} \frac{\Gamma^2}{64}\left[ 4\delta_0(\Delta)^2+5\delta_1(\Delta)^2+7\delta_2(\Delta)^2\right]
\end{align}
which also scales as $\eta_{\gamma} \propto1/\Delta^2$ for large detuning.
$k(\Delta)$ is a correction factor that accounts for the fact that a fraction of the scattering events leaves the state unchanged.
A good metric to compare measurement strategies is the number of photons $\NL$ required to achieve a given sensitivity.
Minimizing this metric will minimize damage to the atomic state independently of the correction factor $k(\Delta)$.
For our calculations, we set $k(\Delta)=0.4$, which predicts our measurements at large detuning

An estimate for the quantum-noise reduction that can be achieved in a single-pass measurement, valid for $\eta_{\rm sc}\ll1$, is given by
\begin{align}
	\label{eq:squeezing}
	\xi^2 &= \frac{1}{1+\zeta}+2\eta_{\rm sc}
\end{align}
where $\zeta$ is the signal-to-noise ratio of the measurement, i.e., the ratio of atomic quantum noise to light shot noise in the measured variance $(\Delta\Sy\supout)^2$.
For the two strategies considered here,
\begin{align}
	\label{eq:snrAOC}
	\zeta_{\rm AOC} &= \frac{\kA^2\NL\NA}{4}\left(1+\frac{\kB^2\NL^2}{16}\right)
\end{align}
and
\begin{align}
	\label{eq:snrLTE}
	\zeta_{\rm \LIN} &= \frac{\kA^2\NL\NA}{4}.
\end{align}
Metrologically significant squeezing is then given by 
\begin{align}
	\label{eq:met}
	\xi_m^2 =\xi^2/(1-\eta_{\rm sc})^2.
\end{align}


\begin{acknowledgments}
We thank C. Caves, I. Walmsely, A. Datta, and J. Nunn for useful discussions, and M. Koschorreck and R. P. Anderson for helpful comments.
This work was supported by the Spanish Ministerio de Econom\'{i}a y Competitividad under the project Magnetometria ultra-precisa basada en optica quantica (MAGO) (Reference No. FIS2011-23520), by the European Research Council  under the project Atomic Quantum Metrology (AQUMET) and by Fundaci\'{o} Privada CELLEX Barcelona.
\end{acknowledgments}



\begin{thebibliography}{62}%
\makeatletter
\providecommand \@ifxundefined [1]{%
 \@ifx{#1\undefined}
}%
\providecommand \@ifnum [1]{%
 \ifnum #1\expandafter \@firstoftwo
 \else \expandafter \@secondoftwo
 \fi
}%
\providecommand \@ifx [1]{%
 \ifx #1\expandafter \@firstoftwo
 \else \expandafter \@secondoftwo
 \fi
}%
\providecommand \natexlab [1]{#1}%
\providecommand \enquote  [1]{{\em #1}}%
\providecommand \bibnamefont  [1]{#1}%
\providecommand \bibfnamefont [1]{#1}%
\providecommand \citenamefont [1]{#1}%
\providecommand \href@noop [0]{\@secondoftwo}%
\providecommand \href [0]{\begingroup \@sanitize@url \@href}%
\providecommand \@href[1]{\@@startlink{#1}\@@href}%
\providecommand \@@href[1]{\endgroup#1\@@endlink}%
\providecommand \@sanitize@url [0]{\catcode `\\12\catcode `\$12\catcode
  `\&12\catcode `\#12\catcode `\^12\catcode `\_12\catcode `\%12\relax}%
\providecommand \@@startlink[1]{}%
\providecommand \@@endlink[0]{}%
\providecommand \url  [0]{\begingroup\@sanitize@url \@url }%
\providecommand \@url [1]{\endgroup\@href {#1}{\urlprefix }}%
\providecommand \urlprefix  [0]{URL }%
\providecommand \Eprint [0]{\href }%
\providecommand \doibase [0]{http://dx.doi.org/}%
\providecommand \selectlanguage [0]{\@gobble}%
\providecommand \bibinfo  [0]{\@secondoftwo}%
\providecommand \bibfield  [0]{\@secondoftwo}%
\providecommand \translation [1]{[#1]}%
\providecommand \BibitemOpen [0]{}%
\providecommand \bibitemStop [0]{}%
\providecommand \bibitemNoStop [0]{.\EOS\space}%
\providecommand \EOS [0]{\spacefactor3000\relax}%
\providecommand \BibitemShut  [1]{\csname bibitem#1\endcsname}%
\let\auto@bib@innerbib\@empty
\bibitem [{\citenamefont {Allred}\ \emph {et~al.}(2002)\citenamefont {Allred},
  \citenamefont {Lyman}, \citenamefont {Kornack},\ and\ \citenamefont
  {Romalis}}]{Allred2002}%
  \BibitemOpen
  \bibfield  {author} {\bibinfo {author} {\bibfnamefont {J.~C.}\ \bibnamefont
  {Allred}}, \bibinfo {author} {\bibfnamefont {R.~N.}\ \bibnamefont {Lyman}},
  \bibinfo {author} {\bibfnamefont {T.~W.}\ \bibnamefont {Kornack}}, \ and\
  \bibinfo {author} {\bibfnamefont {M.~V.}\ \bibnamefont {Romalis}},\
  }\bibfield  {title} {\enquote {\bibinfo {title} {High-sensitivity atomic
  magnetometer unaffected by spin-exchange relaxation},}\ }\href {\doibase
  10.1103/PhysRevLett.89.130801} {\bibfield  {journal} {\bibinfo  {journal}
  {Phys. Rev. Lett.}\ }\textbf {\bibinfo {volume} {89}},\ \bibinfo {pages}
  {130801} (\bibinfo {year} {2002})}\BibitemShut {NoStop}%
\bibitem [{\citenamefont {Budker}\ \emph {et~al.}(2000)\citenamefont {Budker},
  \citenamefont {Kimball}, \citenamefont {Rochester},\ and\ \citenamefont
  {Yashchuk}}]{Budker2000a}%
  \BibitemOpen
  \bibfield  {author} {\bibinfo {author} {\bibfnamefont {D.}~\bibnamefont
  {Budker}}, \bibinfo {author} {\bibfnamefont {D.~F.}\ \bibnamefont {Kimball}},
  \bibinfo {author} {\bibfnamefont {S.~M.}\ \bibnamefont {Rochester}}, \ and\
  \bibinfo {author} {\bibfnamefont {V.~V.}\ \bibnamefont {Yashchuk}},\
  }\bibfield  {title} {\enquote {\bibinfo {title} {Nonlinear magneto-optical
  rotation via alignment-to-orientation conversion},}\ }\href {\doibase
  10.1103/PhysRevLett.85.2088} {\bibfield  {journal} {\bibinfo  {journal}
  {Phys. Rev. Lett.}\ }\textbf {\bibinfo {volume} {85}},\ \bibinfo {pages}
  {2088-2091} (\bibinfo {year} {2000})}\BibitemShut {NoStop}%
\bibitem [{\citenamefont {Schumm}\ \emph {et~al.}(2005)\citenamefont {Schumm},
  \citenamefont {Hofferberth}, \citenamefont {Andersson}, \citenamefont
  {Wildermuth}, \citenamefont {Groth}, \citenamefont {Bar-Joseph},
  \citenamefont {Schmiedmayer},\ and\ \citenamefont
  {Kr{\"u}ger}}]{Schumm2005a}%
  \BibitemOpen
  \bibfield  {author} {\bibinfo {author} {\bibfnamefont {T.}~\bibnamefont
  {Schumm}}, \bibinfo {author} {\bibfnamefont {S.}~\bibnamefont {Hofferberth}},
  \bibinfo {author} {\bibfnamefont {L.~M.}\ \bibnamefont {Andersson}}, \bibinfo
  {author} {\bibfnamefont {S.}~\bibnamefont {Wildermuth}}, \bibinfo {author}
  {\bibfnamefont {S.}~\bibnamefont {Groth}}, \bibinfo {author} {\bibfnamefont
  {I.}~\bibnamefont {Bar-Joseph}}, \bibinfo {author} {\bibfnamefont
  {J.}~\bibnamefont {Schmiedmayer}}, \ and\ \bibinfo {author} {\bibfnamefont
  {P.}~\bibnamefont {Kr{\"u}ger}},\ }\bibfield  {title} {\enquote {\bibinfo
  {title} {Matter-wave interferometry in a double well on an atom chip},}\
  }\href {\doibase 10.1038/nphys125} {\bibfield  {journal} {\bibinfo  {journal}
  {Nat. Phys.}\ }\textbf {\bibinfo {volume} {1}},\ \bibinfo {pages} {57-62}
  (\bibinfo {year} {2005})}\BibitemShut {NoStop}%
\bibitem [{\citenamefont {Jo}\ \emph {et~al.}(2007{\natexlab{a}})\citenamefont
  {Jo}, \citenamefont {Choi}, \citenamefont {Christensen}, \citenamefont {Lee},
  \citenamefont {Pasquini}, \citenamefont {Ketterle},\ and\ \citenamefont
  {Pritchard}}]{Jo2007b}%
  \BibitemOpen
  \bibfield  {author} {\bibinfo {author} {\bibfnamefont {G.-B.}\ \bibnamefont
  {Jo}}, \bibinfo {author} {\bibfnamefont {J.-H.}\ \bibnamefont {Choi}},
  \bibinfo {author} {\bibfnamefont {C.~A.}\ \bibnamefont {Christensen}},
  \bibinfo {author} {\bibfnamefont {Y.-R.}\ \bibnamefont {Lee}}, \bibinfo
  {author} {\bibfnamefont {T.~A.}\ \bibnamefont {Pasquini}}, \bibinfo {author}
  {\bibfnamefont {W.}~\bibnamefont {Ketterle}}, \ and\ \bibinfo {author}
  {\bibfnamefont {D.~E.}\ \bibnamefont {Pritchard}},\ }\bibfield  {title}
  {\enquote {\bibinfo {title} {Matter-wave interferometry with phase
  fluctuating bose-einstein condensates},}\ }\href {\doibase
  10.1103/PhysRevLett.99.240406} {\bibfield  {journal} {\bibinfo  {journal}
  {Phys. Rev. Lett.}\ }\textbf {\bibinfo {volume} {99}},\ \bibinfo {pages}
  {240406} (\bibinfo {year} {2007}{\natexlab{a}})}\BibitemShut {NoStop}%
\bibitem [{\citenamefont {Jo}\ \emph {et~al.}(2007{\natexlab{b}})\citenamefont
  {Jo}, \citenamefont {Choi}, \citenamefont {Christensen}, \citenamefont
  {Pasquini}, \citenamefont {Lee}, \citenamefont {Ketterle},\ and\
  \citenamefont {Pritchard}}]{Jo2007c}%
  \BibitemOpen
  \bibfield  {author} {\bibinfo {author} {\bibfnamefont {G.-B.}\ \bibnamefont
  {Jo}}, \bibinfo {author} {\bibfnamefont {J.-H.}\ \bibnamefont {Choi}},
  \bibinfo {author} {\bibfnamefont {C.~A.}\ \bibnamefont {Christensen}},
  \bibinfo {author} {\bibfnamefont {T.~A.}\ \bibnamefont {Pasquini}}, \bibinfo
  {author} {\bibfnamefont {Y.-R.}\ \bibnamefont {Lee}}, \bibinfo {author}
  {\bibfnamefont {W.}~\bibnamefont {Ketterle}}, \ and\ \bibinfo {author}
  {\bibfnamefont {D.~E.}\ \bibnamefont {Pritchard}},\ }\bibfield  {title}
  {\enquote {\bibinfo {title} {Phase-sensitive recombination of two
  bose-einstein condensates on an atom chip},}\ }\href {\doibase
  10.1103/PhysRevLett.98.180401} {\bibfield  {journal} {\bibinfo  {journal}
  {Phys. Rev. Lett.}\ }\textbf {\bibinfo {volume} {98}},\ \bibinfo {pages}
  {180401} (\bibinfo {year} {2007}{\natexlab{b}})}\BibitemShut {NoStop}%
\bibitem [{\citenamefont {Baumg\"{a}rtner}\ \emph {et~al.}(2010)\citenamefont
  {Baumg\"{a}rtner}, \citenamefont {Sewell}, \citenamefont {Eriksson},
  \citenamefont {Llorente-Garcia}, \citenamefont {Dingjan}, \citenamefont
  {Cotter},\ and\ \citenamefont {Hinds}}]{Baumgaertner2010a}%
  \BibitemOpen
  \bibfield  {author} {\bibinfo {author} {\bibfnamefont {Florian}\ \bibnamefont
  {Baumg\"{a}rtner}}, \bibinfo {author} {\bibfnamefont {R.~J.}\ \bibnamefont
  {Sewell}}, \bibinfo {author} {\bibfnamefont {S.}~\bibnamefont {Eriksson}},
  \bibinfo {author} {\bibfnamefont {I.}~\bibnamefont {Llorente-Garcia}},
  \bibinfo {author} {\bibfnamefont {Jos}\ \bibnamefont {Dingjan}}, \bibinfo
  {author} {\bibfnamefont {J.~P.}\ \bibnamefont {Cotter}}, \ and\ \bibinfo
  {author} {\bibfnamefont {E.~A.}\ \bibnamefont {Hinds}},\ }\bibfield  {title}
  {\enquote {\bibinfo {title} {{Measuring Energy Differences by BEC
  Interferometry on a Chip}},}\ }\href {\doibase
  10.1103/PhysRevLett.105.243003} {\bibfield  {journal} {\bibinfo  {journal}
  {Phys. Rev. Lett.}\ }\textbf {\bibinfo {volume} {105}},\ \bibinfo {pages}
  {243003} (\bibinfo {year} {2010})}\BibitemShut {NoStop}%
\bibitem [{\citenamefont {Wolfgramm}\ \emph {et~al.}(2010)\citenamefont
  {Wolfgramm}, \citenamefont {Cer\`e}, \citenamefont {Beduini}, \citenamefont
  {Predojevi\ifmmode~\acute{c}\else \'{c}\fi{}}, \citenamefont {Koschorreck},\
  and\ \citenamefont {Mitchell}}]{Wolfgramm2010}%
  \BibitemOpen
  \bibfield  {author} {\bibinfo {author} {\bibfnamefont {Florian}\ \bibnamefont
  {Wolfgramm}}, \bibinfo {author} {\bibfnamefont {Alessandro}\ \bibnamefont
  {Cer\`e}}, \bibinfo {author} {\bibfnamefont {Federica~A.}\ \bibnamefont
  {Beduini}}, \bibinfo {author} {\bibfnamefont {Ana}\ \bibnamefont
  {Predojevi\ifmmode~\acute{c}\else \'{c}\fi{}}}, \bibinfo {author}
  {\bibfnamefont {Marco}\ \bibnamefont {Koschorreck}}, \ and\ \bibinfo {author}
  {\bibfnamefont {Morgan~W.}\ \bibnamefont {Mitchell}},\ }\bibfield  {title}
  {\enquote {\bibinfo {title} {Squeezed-light optical magnetometry},}\ }\href
  {\doibase 10.1103/PhysRevLett.105.053601} {\bibfield  {journal} {\bibinfo
  {journal} {Phys. Rev. Lett.}\ }\textbf {\bibinfo {volume} {105}},\ \bibinfo
  {pages} {053601} (\bibinfo {year} {2010})}\BibitemShut {NoStop}%
\bibitem [{\citenamefont {Napolitano}\ \emph {et~al.}(2011)\citenamefont
  {Napolitano}, \citenamefont {Koschorreck}, \citenamefont {Dubost},
  \citenamefont {Behbood}, \citenamefont {Sewell},\ and\ \citenamefont
  {Mitchell}}]{Napolitano2011}%
  \BibitemOpen
  \bibfield  {author} {\bibinfo {author} {\bibfnamefont {M.}~\bibnamefont
  {Napolitano}}, \bibinfo {author} {\bibfnamefont {M.}~\bibnamefont
  {Koschorreck}}, \bibinfo {author} {\bibfnamefont {B.}~\bibnamefont {Dubost}},
  \bibinfo {author} {\bibfnamefont {N.}~\bibnamefont {Behbood}}, \bibinfo
  {author} {\bibfnamefont {R.~J.}\ \bibnamefont {Sewell}}, \ and\ \bibinfo
  {author} {\bibfnamefont {M.~W.}\ \bibnamefont {Mitchell}},\ }\bibfield
  {title} {\enquote {\bibinfo {title} {Interaction-based quantum metrology
  showing scaling beyond the heisenberg limit},}\ }\href {\doibase
  10.1038/nature09778} {\bibfield  {journal} {\bibinfo  {journal} {Nature}\
  }\textbf {\bibinfo {volume} {471}},\ \bibinfo {pages} {486-489} (\bibinfo
  {year} {2011})}\BibitemShut {NoStop}%
\bibitem [{\citenamefont {Horrom}\ \emph {et~al.}(2012)\citenamefont {Horrom},
  \citenamefont {Singh}, \citenamefont {Dowling},\ and\ \citenamefont
  {Mikhailov}}]{Horrom2012}%
  \BibitemOpen
  \bibfield  {author} {\bibinfo {author} {\bibfnamefont {Travis}\ \bibnamefont
  {Horrom}}, \bibinfo {author} {\bibfnamefont {Robinjeet}\ \bibnamefont
  {Singh}}, \bibinfo {author} {\bibfnamefont {Jonathan~P.}\ \bibnamefont
  {Dowling}}, \ and\ \bibinfo {author} {\bibfnamefont {Eugeniy~E.}\
  \bibnamefont {Mikhailov}},\ }\bibfield  {title} {\enquote {\bibinfo {title}
  {Quantum-enhanced magnetometer with low-frequency squeezing},}\ }\href
  {\doibase 10.1103/PhysRevA.86.023803} {\bibfield  {journal} {\bibinfo
  {journal} {Phys. Rev. A}\ }\textbf {\bibinfo {volume} {86}},\ \bibinfo
  {pages} {023803} (\bibinfo {year} {2012})}\BibitemShut {NoStop}%
\bibitem [{\citenamefont {Hamley}\ \emph {et~al.}(2012)\citenamefont {Hamley},
  \citenamefont {Gerving}, \citenamefont {Hoang}, \citenamefont {Bookjans},\
  and\ \citenamefont {Chapman}}]{Hamley2012}%
  \BibitemOpen
  \bibfield  {author} {\bibinfo {author} {\bibfnamefont {C.~D.}\ \bibnamefont
  {Hamley}}, \bibinfo {author} {\bibfnamefont {C.~S.}\ \bibnamefont {Gerving}},
  \bibinfo {author} {\bibfnamefont {T.~M.}\ \bibnamefont {Hoang}}, \bibinfo
  {author} {\bibfnamefont {E.~M.}\ \bibnamefont {Bookjans}}, \ and\ \bibinfo
  {author} {\bibfnamefont {M.~S.}\ \bibnamefont {Chapman}},\ }\bibfield
  {title} {\enquote {\bibinfo {title} {Spin-nematic squeezed vacuum in a
  quantum gas},}\ }\href {\doibase 10.1038/nphys2245} {\bibfield  {journal}
  {\bibinfo  {journal} {Nat. Phys.}\ }\textbf {\bibinfo {volume} {8}},\
  \bibinfo {pages} {305-308} (\bibinfo {year} {2012})}\BibitemShut {NoStop}%
\bibitem [{\citenamefont {Sewell}\ \emph {et~al.}(2012)\citenamefont {Sewell},
  \citenamefont {Koschorreck}, \citenamefont {Napolitano}, \citenamefont
  {Dubost}, \citenamefont {Behbood},\ and\ \citenamefont
  {Mitchell}}]{Sewell2012}%
  \BibitemOpen
  \bibfield  {author} {\bibinfo {author} {\bibfnamefont {R.~J.}\ \bibnamefont
  {Sewell}}, \bibinfo {author} {\bibfnamefont {M.}~\bibnamefont {Koschorreck}},
  \bibinfo {author} {\bibfnamefont {M.}~\bibnamefont {Napolitano}}, \bibinfo
  {author} {\bibfnamefont {B.}~\bibnamefont {Dubost}}, \bibinfo {author}
  {\bibfnamefont {N.}~\bibnamefont {Behbood}}, \ and\ \bibinfo {author}
  {\bibfnamefont {M.~W.}\ \bibnamefont {Mitchell}},\ }\bibfield  {title}
  {\enquote {\bibinfo {title} {Magnetic sensitivity beyond the projection noise
  limit by spin squeezing},}\ }\href {\doibase 10.1103/PhysRevLett.109.253605}
  {\bibfield  {journal} {\bibinfo  {journal} {Phys. Rev. Lett.}\ }\textbf
  {\bibinfo {volume} {109}},\ \bibinfo {pages} {253605} (\bibinfo {year}
  {2012})}\BibitemShut {NoStop}%
\bibitem [{\citenamefont {Wolfgramm}\ \emph {et~al.}(2013)\citenamefont
  {Wolfgramm}, \citenamefont {Vitelli}, \citenamefont {Beduini}, \citenamefont
  {Godbout},\ and\ \citenamefont {Mitchell}}]{WolfgrammNPhot2013}%
  \BibitemOpen
  \bibfield  {author} {\bibinfo {author} {\bibfnamefont {Florian}\ \bibnamefont
  {Wolfgramm}}, \bibinfo {author} {\bibfnamefont {Chiara}\ \bibnamefont
  {Vitelli}}, \bibinfo {author} {\bibfnamefont {Federica~A.}\ \bibnamefont
  {Beduini}}, \bibinfo {author} {\bibfnamefont {Nicolas}\ \bibnamefont
  {Godbout}}, \ and\ \bibinfo {author} {\bibfnamefont {Morgan~W.}\ \bibnamefont
  {Mitchell}},\ }\bibfield  {title} {\enquote {\bibinfo {title}
  {Entanglement-enhanced probing of a delicate material system},}\ }\href
  {\doibase 10.1038/nphoton.2012.300} {\bibfield  {journal} {\bibinfo
  {journal} {Nat. Photon.}\ }\textbf {\bibinfo {volume} {7}},\ \bibinfo {pages}
  {28-32} (\bibinfo {year} {2013})}\BibitemShut {NoStop}%
\bibitem [{\citenamefont {Esteve}\ \emph {et~al.}(2008)\citenamefont {Esteve},
  \citenamefont {Gross}, \citenamefont {Weller}, \citenamefont {Giovanazzi},\
  and\ \citenamefont {Oberthaler}}]{Esteve2008}%
  \BibitemOpen
  \bibfield  {author} {\bibinfo {author} {\bibfnamefont {J.}~\bibnamefont
  {Esteve}}, \bibinfo {author} {\bibfnamefont {C.}~\bibnamefont {Gross}},
  \bibinfo {author} {\bibfnamefont {A.}~\bibnamefont {Weller}}, \bibinfo
  {author} {\bibfnamefont {S.}~\bibnamefont {Giovanazzi}}, \ and\ \bibinfo
  {author} {\bibfnamefont {M.~K.}\ \bibnamefont {Oberthaler}},\ }\bibfield
  {title} {\enquote {\bibinfo {title} {Squeezing and entanglement in a
  bose-einstein condensate},}\ }\href {\doibase 10.1038/nature07332} {\bibfield
   {journal} {\bibinfo  {journal} {Nature}\ }\textbf {\bibinfo {volume}
  {455}},\ \bibinfo {pages} {1216-1219} (\bibinfo {year} {2008})}\BibitemShut
  {NoStop}%
\bibitem [{\citenamefont {Riedel}\ \emph {et~al.}(2010)\citenamefont {Riedel},
  \citenamefont {B\"{o}hl}, \citenamefont {Li}, \citenamefont {H\"{a}nsch},
  \citenamefont {Sinatra},\ and\ \citenamefont {Treutlein}}]{Riedel2010}%
  \BibitemOpen
  \bibfield  {author} {\bibinfo {author} {\bibfnamefont {Max~F.}\ \bibnamefont
  {Riedel}}, \bibinfo {author} {\bibfnamefont {Pascal}\ \bibnamefont
  {B\"{o}hl}}, \bibinfo {author} {\bibfnamefont {Yun}\ \bibnamefont {Li}},
  \bibinfo {author} {\bibfnamefont {Theodor~W.}\ \bibnamefont {H\"{a}nsch}},
  \bibinfo {author} {\bibfnamefont {Alice}\ \bibnamefont {Sinatra}}, \ and\
  \bibinfo {author} {\bibfnamefont {Philipp}\ \bibnamefont {Treutlein}},\
  }\bibfield  {title} {\enquote {\bibinfo {title} {Atom-chip-based generation
  of entanglement for quantum metrology},}\ }\href {\doibase
  10.1038/nature08988} {\bibfield  {journal} {\bibinfo  {journal} {Nature}\
  }\textbf {\bibinfo {volume} {464}},\ \bibinfo {pages} {1170-1173} (\bibinfo
  {year} {2010})}\BibitemShut {NoStop}%
\bibitem [{\citenamefont {Gross}\ \emph {et~al.}(2010)\citenamefont {Gross},
  \citenamefont {Zibold}, \citenamefont {Nicklas}, \citenamefont {Est\`{e}ve},\
  and\ \citenamefont {Oberthaler}}]{Gross2010}%
  \BibitemOpen
  \bibfield  {author} {\bibinfo {author} {\bibfnamefont {C.}~\bibnamefont
  {Gross}}, \bibinfo {author} {\bibfnamefont {T.}~\bibnamefont {Zibold}},
  \bibinfo {author} {\bibfnamefont {E.}~\bibnamefont {Nicklas}}, \bibinfo
  {author} {\bibfnamefont {J.}~\bibnamefont {Est\`{e}ve}}, \ and\ \bibinfo
  {author} {\bibfnamefont {M.~K.}\ \bibnamefont {Oberthaler}},\ }\bibfield
  {title} {\enquote {\bibinfo {title} {Nonlinear atom interferometer surpasses
  classical precision limit},}\ }\href {\doibase 10.1038/nature08919}
  {\bibfield  {journal} {\bibinfo  {journal} {Nature}\ }\textbf {\bibinfo
  {volume} {464}},\ \bibinfo {pages} {1165-1169} (\bibinfo {year}
  {2010})}\BibitemShut {NoStop}%
\bibitem [{\citenamefont {Gross}\ \emph {et~al.}(2011)\citenamefont {Gross},
  \citenamefont {Strobel}, \citenamefont {Nicklas}, \citenamefont {Zibold},
  \citenamefont {Bar-Gill}, \citenamefont {Kurizki},\ and\ \citenamefont
  {Oberthaler}}]{Gross2011a}%
  \BibitemOpen
  \bibfield  {author} {\bibinfo {author} {\bibfnamefont {C.}~\bibnamefont
  {Gross}}, \bibinfo {author} {\bibfnamefont {H.}~\bibnamefont {Strobel}},
  \bibinfo {author} {\bibfnamefont {E.}~\bibnamefont {Nicklas}}, \bibinfo
  {author} {\bibfnamefont {T.}~\bibnamefont {Zibold}}, \bibinfo {author}
  {\bibfnamefont {N.}~\bibnamefont {Bar-Gill}}, \bibinfo {author}
  {\bibfnamefont {G.}~\bibnamefont {Kurizki}}, \ and\ \bibinfo {author}
  {\bibfnamefont {M.~K.}\ \bibnamefont {Oberthaler}},\ }\bibfield  {title}
  {\enquote {\bibinfo {title} {{Atomic homodyne detection of
  continuous-variable entangled twin-atom states}},}\ }\href {\doibase
  http://dx.doi.org/10.1038/nature10654} {\bibfield  {journal} {\bibinfo
  {journal} {Nature}\ }\textbf {\bibinfo {volume} {480}},\ \bibinfo {pages}
  {219} (\bibinfo {year} {2011})}\BibitemShut {NoStop}%
\bibitem [{\citenamefont {Luecke}\ \emph {et~al.}(2011)\citenamefont {Luecke},
  \citenamefont {Scherer}, \citenamefont {Kruse}, \citenamefont {Pezze},
  \citenamefont {Deuretzbacher}, \citenamefont {Hyllus}, \citenamefont {Topic},
  \citenamefont {Peise}, \citenamefont {Ertmer}, \citenamefont {Arlt},
  \citenamefont {Santos}, \citenamefont {Smerzi},\ and\ \citenamefont
  {Klempt}}]{Luecke2011}%
  \BibitemOpen
  \bibfield  {author} {\bibinfo {author} {\bibfnamefont {B.}~\bibnamefont
  {Luecke}}, \bibinfo {author} {\bibfnamefont {M.}~\bibnamefont {Scherer}},
  \bibinfo {author} {\bibfnamefont {J.}~\bibnamefont {Kruse}}, \bibinfo
  {author} {\bibfnamefont {L.}~\bibnamefont {Pezze}}, \bibinfo {author}
  {\bibfnamefont {F.}~\bibnamefont {Deuretzbacher}}, \bibinfo {author}
  {\bibfnamefont {P.}~\bibnamefont {Hyllus}}, \bibinfo {author} {\bibfnamefont
  {O.}~\bibnamefont {Topic}}, \bibinfo {author} {\bibfnamefont
  {J.}~\bibnamefont {Peise}}, \bibinfo {author} {\bibfnamefont
  {W.}~\bibnamefont {Ertmer}}, \bibinfo {author} {\bibfnamefont
  {J.}~\bibnamefont {Arlt}}, \bibinfo {author} {\bibfnamefont {L.}~\bibnamefont
  {Santos}}, \bibinfo {author} {\bibfnamefont {A.}~\bibnamefont {Smerzi}}, \
  and\ \bibinfo {author} {\bibfnamefont {C.}~\bibnamefont {Klempt}},\
  }\bibfield  {title} {\enquote {\bibinfo {title} {Twin matter waves for
  interferometry beyond the classical limit},}\ }\href {\doibase
  10.1126/science.1208798} {\bibfield  {journal} {\bibinfo  {journal}
  {Science}\ }\textbf {\bibinfo {volume} {334}},\ \bibinfo {pages} {773-776}
  (\bibinfo {year} {2011})}\BibitemShut {NoStop}%
\bibitem [{\citenamefont {B{\"u}cker}\ \emph {et~al.}(2011)\citenamefont
  {B{\"u}cker}, \citenamefont {Grond}, \citenamefont {Manz}, \citenamefont
  {Berrada}, \citenamefont {Betz}, \citenamefont {Koller}, \citenamefont
  {Hohenester}, \citenamefont {Schumm}, \citenamefont {Perrin},\ and\
  \citenamefont {Schmiedmayer}}]{Bucker2011a}%
  \BibitemOpen
  \bibfield  {author} {\bibinfo {author} {\bibfnamefont {Robert}\ \bibnamefont
  {B{\"u}cker}}, \bibinfo {author} {\bibfnamefont {Julian}\ \bibnamefont
  {Grond}}, \bibinfo {author} {\bibfnamefont {Stephanie}\ \bibnamefont {Manz}},
  \bibinfo {author} {\bibfnamefont {Tarik}\ \bibnamefont {Berrada}}, \bibinfo
  {author} {\bibfnamefont {Thomas}\ \bibnamefont {Betz}}, \bibinfo {author}
  {\bibfnamefont {Christian}\ \bibnamefont {Koller}}, \bibinfo {author}
  {\bibfnamefont {Ulrich}\ \bibnamefont {Hohenester}}, \bibinfo {author}
  {\bibfnamefont {Thorsten}\ \bibnamefont {Schumm}}, \bibinfo {author}
  {\bibfnamefont {Aur{\'e}lien}\ \bibnamefont {Perrin}}, \ and\ \bibinfo
  {author} {\bibfnamefont {J{\"o}rg}\ \bibnamefont {Schmiedmayer}},\ }\bibfield
   {title} {\enquote {\bibinfo {title} {{Twin-atom beams}},}\ }\href {\doibase
  10.1038/NPHYS1992} {\bibfield  {journal} {\bibinfo  {journal} {Nat. Phys.}\
  }\textbf {\bibinfo {volume} {7}},\ \bibinfo {pages} {608-611} (\bibinfo
  {year} {2011})}\BibitemShut {NoStop}%
\bibitem [{\citenamefont {Brahms}\ \emph {et~al.}(2011)\citenamefont {Brahms},
  \citenamefont {Purdy}, \citenamefont {Brooks}, \citenamefont {Botter},\ and\
  \citenamefont {Stamper-Kurn}}]{Brahms2011}%
  \BibitemOpen
  \bibfield  {author} {\bibinfo {author} {\bibfnamefont {Nathan}\ \bibnamefont
  {Brahms}}, \bibinfo {author} {\bibfnamefont {Thomas~P}\ \bibnamefont
  {Purdy}}, \bibinfo {author} {\bibfnamefont {Daniel W~C}\ \bibnamefont
  {Brooks}}, \bibinfo {author} {\bibfnamefont {Thierry}\ \bibnamefont
  {Botter}}, \ and\ \bibinfo {author} {\bibfnamefont {Dan~M}\ \bibnamefont
  {Stamper-Kurn}},\ }\bibfield  {title} {\enquote {\bibinfo {title}
  {{Cavity-aided magnetic resonance microscopy of atomic transport in optical
  lattices}},}\ }\href {\doibase 10.1038/nphys1967} {\bibfield  {journal}
  {\bibinfo  {journal} {Nat. Phys.}\ }\textbf {\bibinfo {volume} {7}},\
  \bibinfo {pages} {604-607} (\bibinfo {year} {2011})}\BibitemShut {NoStop}%
\bibitem [{\citenamefont {Berrada}\ \emph {et~al.}(2013)\citenamefont
  {Berrada}, \citenamefont {van Frank}, \citenamefont {B{\"u}cker},
  \citenamefont {Schumm}, \citenamefont {Schaff},\ and\ \citenamefont
  {Schmiedmayer}}]{Berrada2013a}%
  \BibitemOpen
  \bibfield  {author} {\bibinfo {author} {\bibfnamefont {T.}~\bibnamefont
  {Berrada}}, \bibinfo {author} {\bibfnamefont {S.}~\bibnamefont {van Frank}},
  \bibinfo {author} {\bibfnamefont {R.}~\bibnamefont {B{\"u}cker}}, \bibinfo
  {author} {\bibfnamefont {T.}~\bibnamefont {Schumm}}, \bibinfo {author}
  {\bibfnamefont {J.~F.}\ \bibnamefont {Schaff}}, \ and\ \bibinfo {author}
  {\bibfnamefont {J.}~\bibnamefont {Schmiedmayer}},\ }\bibfield  {title}
  {\enquote {\bibinfo {title} {{Integrated Mach-Zehnder interferometer for
  Bose-Einstein condensates}},}\ }\href {\doibase 10.1038/ncomms3077}
  {\bibfield  {journal} {\bibinfo  {journal} {Nat. Comm.}\ }\textbf {\bibinfo
  {volume} {4}},\ \bibinfo {pages} {2077} (\bibinfo {year} {2013})}\BibitemShut
  {NoStop}%
\bibitem [{\citenamefont {Caves}(1981)}]{Caves1981}%
  \BibitemOpen
  \bibfield  {author} {\bibinfo {author} {\bibfnamefont {Carlton~M.}\
  \bibnamefont {Caves}},\ }\bibfield  {title} {\enquote {\bibinfo {title}
  {Quantum-mechanical noise in an interferometer},}\ }\href {\doibase
  10.1103/PhysRevD.23.1693} {\bibfield  {journal} {\bibinfo  {journal} {Phys.
  Rev. D}\ }\textbf {\bibinfo {volume} {23}},\ \bibinfo {pages} {1693-1708}
  (\bibinfo {year} {1981})}\BibitemShut {NoStop}%
\bibitem [{\citenamefont {Giovannetti}\ \emph {et~al.}(2004)\citenamefont
  {Giovannetti}, \citenamefont {Lloyd},\ and\ \citenamefont
  {Maccone}}]{Giovannetti2004}%
  \BibitemOpen
  \bibfield  {author} {\bibinfo {author} {\bibfnamefont {Vittorio}\
  \bibnamefont {Giovannetti}}, \bibinfo {author} {\bibfnamefont {Seth}\
  \bibnamefont {Lloyd}}, \ and\ \bibinfo {author} {\bibfnamefont {Lorenzo}\
  \bibnamefont {Maccone}},\ }\bibfield  {title} {\enquote {\bibinfo {title}
  {Quantum-enhanced measurements: Beating the standard quantum limit},}\ }\href
  {\doibase 10.1126/science.1104149} {\bibfield  {journal} {\bibinfo  {journal}
  {Science}\ }\textbf {\bibinfo {volume} {306}},\ \bibinfo {pages} {1330-1336}
  (\bibinfo {year} {2004})}\BibitemShut {NoStop}%
\bibitem [{\citenamefont {Giovannetti}\ \emph {et~al.}(2006)\citenamefont
  {Giovannetti}, \citenamefont {Lloyd},\ and\ \citenamefont
  {Maccone}}]{Giovannetti2006}%
  \BibitemOpen
  \bibfield  {author} {\bibinfo {author} {\bibfnamefont {Vittorio}\
  \bibnamefont {Giovannetti}}, \bibinfo {author} {\bibfnamefont {Seth}\
  \bibnamefont {Lloyd}}, \ and\ \bibinfo {author} {\bibfnamefont {Lorenzo}\
  \bibnamefont {Maccone}},\ }\bibfield  {title} {\enquote {\bibinfo {title}
  {Quantum metrology},}\ }\href {\doibase 10.1103/PhysRevLett.96.010401}
  {\bibfield  {journal} {\bibinfo  {journal} {Phys. Rev. Lett.}\ }\textbf
  {\bibinfo {volume} {96}},\ \bibinfo {pages} {010401} (\bibinfo {year}
  {2006})}\BibitemShut {NoStop}%
\bibitem [{\citenamefont {Giovannetti}\ \emph {et~al.}(2011)\citenamefont
  {Giovannetti}, \citenamefont {Lloyd},\ and\ \citenamefont
  {Maccone}}]{Giovannetti2011}%
  \BibitemOpen
  \bibfield  {author} {\bibinfo {author} {\bibfnamefont {Vittorio}\
  \bibnamefont {Giovannetti}}, \bibinfo {author} {\bibfnamefont {Seth}\
  \bibnamefont {Lloyd}}, \ and\ \bibinfo {author} {\bibfnamefont {Lorenzo}\
  \bibnamefont {Maccone}},\ }\bibfield  {title} {\enquote {\bibinfo {title}
  {Advances in quantum metrology},}\ }\href {\doibase doi:
  10.1038/nphoton.2011.35} {\bibfield  {journal} {\bibinfo  {journal} {Nat.
  Photon.}\ }\textbf {\bibinfo {volume} {5}},\ \bibinfo {pages} {222-9}
  (\bibinfo {year} {2011})}\BibitemShut {NoStop}%
\bibitem [{\citenamefont {Boixo}\ \emph {et~al.}(2008)\citenamefont {Boixo},
  \citenamefont {Datta}, \citenamefont {Davis}, \citenamefont {Flammia},
  \citenamefont {Shaji},\ and\ \citenamefont {Caves}}]{Boixo2008a}%
  \BibitemOpen
  \bibfield  {author} {\bibinfo {author} {\bibfnamefont {Sergio}\ \bibnamefont
  {Boixo}}, \bibinfo {author} {\bibfnamefont {Animesh}\ \bibnamefont {Datta}},
  \bibinfo {author} {\bibfnamefont {Matthew~J.}\ \bibnamefont {Davis}},
  \bibinfo {author} {\bibfnamefont {Steven~T.}\ \bibnamefont {Flammia}},
  \bibinfo {author} {\bibfnamefont {Anil}\ \bibnamefont {Shaji}}, \ and\
  \bibinfo {author} {\bibfnamefont {Carlton~M.}\ \bibnamefont {Caves}},\
  }\bibfield  {title} {\enquote {\bibinfo {title} {Quantum metrology: Dynamics
  versus entanglement},}\ }\href {\doibase 10.1103/PhysRevLett.101.040403}
  {\bibfield  {journal} {\bibinfo  {journal} {Phys. Rev. Lett.}\ }\textbf
  {\bibinfo {volume} {101}},\ \bibinfo {pages} {040403} (\bibinfo {year}
  {2008})}\BibitemShut {NoStop}%
\bibitem [{\citenamefont {Napolitano}\ and\ \citenamefont
  {Mitchell}(2010)}]{NapolitanoNJP2010}%
  \BibitemOpen
  \bibfield  {author} {\bibinfo {author} {\bibfnamefont {M.}~\bibnamefont
  {Napolitano}}\ and\ \bibinfo {author} {\bibfnamefont {M.~W.}\ \bibnamefont
  {Mitchell}},\ }\bibfield  {title} {\enquote {\bibinfo {title} {Nonlinear
  metrology with a quantum interface},}\ }\href {\doibase
  10.1088/1367-2630/12/9/093016} {\bibfield  {journal} {\bibinfo  {journal}
  {New J. Phys.}\ }\textbf {\bibinfo {volume} {12}},\ \bibinfo {pages} {093016}
  (\bibinfo {year} {2010})}\BibitemShut {NoStop}%
\bibitem [{\citenamefont {Luis}(2004)}]{Luis2004}%
  \BibitemOpen
  \bibfield  {author} {\bibinfo {author} {\bibfnamefont {Alfredo}\ \bibnamefont
  {Luis}},\ }\bibfield  {title} {\enquote {\bibinfo {title} {Nonlinear
  transformations and the heisenberg limit},}\ }\href {\doibase
  http://dx.doi.org/10.1016/j.physleta.2004.06.080} {\bibfield  {journal}
  {\bibinfo  {journal} {Phys. Lett. A}\ }\textbf {\bibinfo {volume} {329}},\
  \bibinfo {pages} {8 - 13} (\bibinfo {year} {2004})}\BibitemShut {NoStop}%
\bibitem [{\citenamefont {Luis}(2007)}]{Luis2007}%
  \BibitemOpen
  \bibfield  {author} {\bibinfo {author} {\bibfnamefont {Alfredo}\ \bibnamefont
  {Luis}},\ }\bibfield  {title} {\enquote {\bibinfo {title} {Quantum limits,
  nonseparable transformations, and nonlinear optics},}\ }\href {\doibase
  10.1103/PhysRevA.76.035801} {\bibfield  {journal} {\bibinfo  {journal} {Phys.
  Rev. A}\ }\textbf {\bibinfo {volume} {76}},\ \bibinfo {eid} {035801}
  (\bibinfo {year} {2007})}\BibitemShut {NoStop}%
\bibitem [{\citenamefont {Rey}\ \emph {et~al.}(2007)\citenamefont {Rey},
  \citenamefont {Jiang},\ and\ \citenamefont {Lukin}}]{Rey2007a}%
  \BibitemOpen
  \bibfield  {author} {\bibinfo {author} {\bibfnamefont {A.~M.}\ \bibnamefont
  {Rey}}, \bibinfo {author} {\bibfnamefont {L.}~\bibnamefont {Jiang}}, \ and\
  \bibinfo {author} {\bibfnamefont {M.~D.}\ \bibnamefont {Lukin}},\ }\bibfield
  {title} {\enquote {\bibinfo {title} {Quantum-limited measurements of atomic
  scattering properties},}\ }\href {\doibase 10.1103/PhysRevA.76.053617}
  {\bibfield  {journal} {\bibinfo  {journal} {Phys. Rev. A}\ }\textbf {\bibinfo
  {volume} {76}},\ \bibinfo {pages} {053617} (\bibinfo {year}
  {2007})}\BibitemShut {NoStop}%
\bibitem [{\citenamefont {Roy}\ and\ \citenamefont
  {Braunstein}(2008)}]{Roy2008}%
  \BibitemOpen
  \bibfield  {author} {\bibinfo {author} {\bibfnamefont {S.~M.}\ \bibnamefont
  {Roy}}\ and\ \bibinfo {author} {\bibfnamefont {Samuel~L.}\ \bibnamefont
  {Braunstein}},\ }\bibfield  {title} {\enquote {\bibinfo {title}
  {Exponentially enhanced quantum metrology},}\ }\href {\doibase
  10.1103/PhysRevLett.100.220501} {\bibfield  {journal} {\bibinfo  {journal}
  {Phys. Rev. Lett.}\ }\textbf {\bibinfo {volume} {100}},\ \bibinfo {pages}
  {220501} (\bibinfo {year} {2008})}\BibitemShut {NoStop}%
\bibitem [{\citenamefont {Choi}\ and\ \citenamefont
  {Sundaram}(2008)}]{Choi2008}%
  \BibitemOpen
  \bibfield  {author} {\bibinfo {author} {\bibfnamefont {S.}~\bibnamefont
  {Choi}}\ and\ \bibinfo {author} {\bibfnamefont {B.}~\bibnamefont
  {Sundaram}},\ }\bibfield  {title} {\enquote {\bibinfo {title} {Bose-einstein
  condensate as a nonlinear ramsey interferometer operating beyond the
  heisenberg limit},}\ }\href {\doibase 10.1103/PhysRevA.77.053613} {\bibfield
  {journal} {\bibinfo  {journal} {Phys. Rev. A}\ }\textbf {\bibinfo {volume}
  {77}},\ \bibinfo {pages} {053613} (\bibinfo {year} {2008})}\BibitemShut
  {NoStop}%
\bibitem [{\citenamefont {Woolley}\ \emph {et~al.}(2008)\citenamefont
  {Woolley}, \citenamefont {Milburn},\ and\ \citenamefont
  {Caves}}]{Woolley2008}%
  \BibitemOpen
  \bibfield  {author} {\bibinfo {author} {\bibfnamefont {M.~J.}\ \bibnamefont
  {Woolley}}, \bibinfo {author} {\bibfnamefont {G.~J.}\ \bibnamefont
  {Milburn}}, \ and\ \bibinfo {author} {\bibfnamefont {Carlton~M.}\
  \bibnamefont {Caves}},\ }\bibfield  {title} {\enquote {\bibinfo {title}
  {Nonlinear quantum metrology using coupled nanomechanical resonators},}\
  }\href {\doibase 10.1088/1367-2630/10/12/125018} {\bibfield  {journal}
  {\bibinfo  {journal} {New J. Phys.}\ }\textbf {\bibinfo {volume} {10}},\
  \bibinfo {pages} {125018} (\bibinfo {year} {2008})}\BibitemShut {NoStop}%
\bibitem [{\citenamefont {Boixo}\ \emph {et~al.}({2009})\citenamefont {Boixo},
  \citenamefont {Datta}, \citenamefont {Davis}, \citenamefont {Shaji},
  \citenamefont {Tacla},\ and\ \citenamefont {Caves}}]{Boixo2009}%
  \BibitemOpen
  \bibfield  {author} {\bibinfo {author} {\bibfnamefont {Sergio}\ \bibnamefont
  {Boixo}}, \bibinfo {author} {\bibfnamefont {Animesh}\ \bibnamefont {Datta}},
  \bibinfo {author} {\bibfnamefont {Matthew~J.}\ \bibnamefont {Davis}},
  \bibinfo {author} {\bibfnamefont {Anil}\ \bibnamefont {Shaji}}, \bibinfo
  {author} {\bibfnamefont {Alexandre~B.}\ \bibnamefont {Tacla}}, \ and\
  \bibinfo {author} {\bibfnamefont {Carlton~M.}\ \bibnamefont {Caves}},\
  }\bibfield  {title} {\enquote {\bibinfo {title} {{Quantum-limited metrology
  and Bose-Einstein condensates}},}\ }\href {\doibase
  {10.1103/PhysRevA.80.032103}} {\bibfield  {journal} {\bibinfo  {journal}
  {Phys. Rev. A}\ }\textbf {\bibinfo {volume} {{80}}},\ \bibinfo {pages}
  {{032103}} (\bibinfo {year} {{2009}})}\BibitemShut {NoStop}%
\bibitem [{\citenamefont {Chase}\ \emph {et~al.}({2009})\citenamefont {Chase},
  \citenamefont {Baragiola}, \citenamefont {Partner}, \citenamefont {Black},\
  and\ \citenamefont {Geremia}}]{Chase2009}%
  \BibitemOpen
  \bibfield  {author} {\bibinfo {author} {\bibfnamefont {Bradley~A.}\
  \bibnamefont {Chase}}, \bibinfo {author} {\bibfnamefont {Ben~Q.}\
  \bibnamefont {Baragiola}}, \bibinfo {author} {\bibfnamefont {Heather~L.}\
  \bibnamefont {Partner}}, \bibinfo {author} {\bibfnamefont {Brigette~D.}\
  \bibnamefont {Black}}, \ and\ \bibinfo {author} {\bibfnamefont {J.~M.}\
  \bibnamefont {Geremia}},\ }\bibfield  {title} {\enquote {\bibinfo {title}
  {{Magnetometry via a double-pass continuous quantum measurement of atomic
  spin}},}\ }\href {\doibase {10.1103/PhysRevA.79.062107}} {\bibfield
  {journal} {\bibinfo  {journal} {Phys. Rev. A}\ }\textbf {\bibinfo {volume}
  {{79}}},\ \bibinfo {pages} {{062107}} (\bibinfo {year} {{2009}})}\BibitemShut
  {NoStop}%
\bibitem [{\citenamefont {Tacla}\ \emph {et~al.}(2010)\citenamefont {Tacla},
  \citenamefont {Boixo}, \citenamefont {Datta}, \citenamefont {Shaji},\ and\
  \citenamefont {Caves}}]{Tacla2010}%
  \BibitemOpen
  \bibfield  {author} {\bibinfo {author} {\bibfnamefont {Alexandre~B.}\
  \bibnamefont {Tacla}}, \bibinfo {author} {\bibfnamefont {Sergio}\
  \bibnamefont {Boixo}}, \bibinfo {author} {\bibfnamefont {Animesh}\
  \bibnamefont {Datta}}, \bibinfo {author} {\bibfnamefont {Anil}\ \bibnamefont
  {Shaji}}, \ and\ \bibinfo {author} {\bibfnamefont {Carlton~M.}\ \bibnamefont
  {Caves}},\ }\bibfield  {title} {\enquote {\bibinfo {title} {Nonlinear
  interferometry with bose-einstein condensates},}\ }\href {\doibase
  10.1103/PhysRevA.82.053636} {\bibfield  {journal} {\bibinfo  {journal} {Phys.
  Rev. A}\ }\textbf {\bibinfo {volume} {82}},\ \bibinfo {pages} {053636}
  (\bibinfo {year} {2010})}\BibitemShut {NoStop}%
\bibitem [{\citenamefont {Tiesinga}\ and\ \citenamefont
  {Johnson}(2013)}]{Tiesinga2013a}%
  \BibitemOpen
  \bibfield  {author} {\bibinfo {author} {\bibfnamefont {E.}~\bibnamefont
  {Tiesinga}}\ and\ \bibinfo {author} {\bibfnamefont {P.~R.}\ \bibnamefont
  {Johnson}},\ }\bibfield  {title} {\enquote {\bibinfo {title} {Quadrature
  interferometry for nonequilibrium ultracold atoms in optical lattices},}\
  }\href {\doibase 10.1103/PhysRevA.87.013423} {\bibfield  {journal} {\bibinfo
  {journal} {Phys. Rev. A}\ }\textbf {\bibinfo {volume} {87}},\ \bibinfo
  {pages} {013423} (\bibinfo {year} {2013})}\BibitemShut {NoStop}%
\bibitem [{\citenamefont {Javanainen}\ and\ \citenamefont
  {Chen}(2012)}]{JavanainenPRA2012}%
  \BibitemOpen
  \bibfield  {author} {\bibinfo {author} {\bibfnamefont {Juha}\ \bibnamefont
  {Javanainen}}\ and\ \bibinfo {author} {\bibfnamefont {Han}\ \bibnamefont
  {Chen}},\ }\bibfield  {title} {\enquote {\bibinfo {title} {Optimal
  measurement precision of a nonlinear interferometer},}\ }\href {\doibase
  10.1103/PhysRevA.85.063605} {\bibfield  {journal} {\bibinfo  {journal} {Phys.
  Rev. A}\ }\textbf {\bibinfo {volume} {85}},\ \bibinfo {pages} {063605}
  (\bibinfo {year} {2012})}\BibitemShut {NoStop}%
\bibitem [{\citenamefont {Zwierz}\ \emph {et~al.}(2010)\citenamefont {Zwierz},
  \citenamefont {P\'erez-Delgado},\ and\ \citenamefont {Kok}}]{Zwierz2010}%
  \BibitemOpen
  \bibfield  {author} {\bibinfo {author} {\bibfnamefont {Marcin}\ \bibnamefont
  {Zwierz}}, \bibinfo {author} {\bibfnamefont {Carlos~A.}\ \bibnamefont
  {P\'erez-Delgado}}, \ and\ \bibinfo {author} {\bibfnamefont {Pieter}\
  \bibnamefont {Kok}},\ }\bibfield  {title} {\enquote {\bibinfo {title}
  {General optimality of the heisenberg limit for quantum metrology},}\ }\href
  {\doibase 10.1103/PhysRevLett.105.180402} {\bibfield  {journal} {\bibinfo
  {journal} {Phys. Rev. Lett.}\ }\textbf {\bibinfo {volume} {105}},\ \bibinfo
  {pages} {180402} (\bibinfo {year} {2010})}\BibitemShut {NoStop}%
\bibitem [{\citenamefont {Demkowicz-Dobrzanski}\ \emph
  {et~al.}(2012)\citenamefont {Demkowicz-Dobrzanski}, \citenamefont
  {Kolodynski},\ and\ \citenamefont {Guta}}]{Demkowicz-Dobrzanski2012}%
  \BibitemOpen
  \bibfield  {author} {\bibinfo {author} {\bibfnamefont {Rafal}\ \bibnamefont
  {Demkowicz-Dobrzanski}}, \bibinfo {author} {\bibfnamefont {Jan}\ \bibnamefont
  {Kolodynski}}, \ and\ \bibinfo {author} {\bibfnamefont {Madalin}\
  \bibnamefont {Guta}},\ }\bibfield  {title} {\enquote {\bibinfo {title} {The
  elusive heisenberg limit in quantum-enhanced metrology},}\ }\href {\doibase
  10.1038/ncomms2067} {\bibfield  {journal} {\bibinfo  {journal} {Nat. Comm.}\
  }\textbf {\bibinfo {volume} {3}},\ \bibinfo {pages} {1063} (\bibinfo {year}
  {2012})}\BibitemShut {NoStop}%
\bibitem [{\citenamefont {Hall}\ and\ \citenamefont
  {Wiseman}(2012)}]{Hall2012a}%
  \BibitemOpen
  \bibfield  {author} {\bibinfo {author} {\bibfnamefont {Michael J.~W.}\
  \bibnamefont {Hall}}\ and\ \bibinfo {author} {\bibfnamefont {Howard~M.}\
  \bibnamefont {Wiseman}},\ }\bibfield  {title} {\enquote {\bibinfo {title}
  {Does nonlinear metrology offer improved resolution? answers from quantum
  information theory},}\ }\href {\doibase 10.1103/PhysRevX.2.041006} {\bibfield
   {journal} {\bibinfo  {journal} {Phys. Rev. X}\ }\textbf {\bibinfo {volume}
  {2}},\ \bibinfo {pages} {041006} (\bibinfo {year} {2012})}\BibitemShut
  {NoStop}%
\bibitem [{\citenamefont {Koschorreck}\ \emph {et~al.}(2010)\citenamefont
  {Koschorreck}, \citenamefont {Napolitano}, \citenamefont {Dubost},\ and\
  \citenamefont {Mitchell}}]{Koschorreck2010b}%
  \BibitemOpen
  \bibfield  {author} {\bibinfo {author} {\bibfnamefont {M.}~\bibnamefont
  {Koschorreck}}, \bibinfo {author} {\bibfnamefont {M.}~\bibnamefont
  {Napolitano}}, \bibinfo {author} {\bibfnamefont {B.}~\bibnamefont {Dubost}},
  \ and\ \bibinfo {author} {\bibfnamefont {M.~W.}\ \bibnamefont {Mitchell}},\
  }\bibfield  {title} {\enquote {\bibinfo {title} {Quantum nondemolition
  measurement of large-spin ensembles by dynamical decoupling},}\ }\href
  {\doibase 10.1103/PhysRevLett.105.093602} {\bibfield  {journal} {\bibinfo
  {journal} {Phys. Rev. Lett.}\ }\textbf {\bibinfo {volume} {105}},\ \bibinfo
  {pages} {093602} (\bibinfo {year} {2010})}\BibitemShut {NoStop}%
\bibitem [{\citenamefont {Sewell}\ \emph {et~al.}(2013)\citenamefont {Sewell},
  \citenamefont {Napolitano}, \citenamefont {Behbood}, \citenamefont
  {Colangelo},\ and\ \citenamefont {Mitchell}}]{SewellNatPhot2013}%
  \BibitemOpen
  \bibfield  {author} {\bibinfo {author} {\bibfnamefont {R.~J.}\ \bibnamefont
  {Sewell}}, \bibinfo {author} {\bibfnamefont {M.}~\bibnamefont {Napolitano}},
  \bibinfo {author} {\bibfnamefont {N.}~\bibnamefont {Behbood}}, \bibinfo
  {author} {\bibfnamefont {G.}~\bibnamefont {Colangelo}}, \ and\ \bibinfo
  {author} {\bibfnamefont {M.~W.}\ \bibnamefont {Mitchell}},\ }\bibfield
  {title} {\enquote {\bibinfo {title} {Certified quantum nondemolition
  measurement of a macroscopic material system},}\ }\href {\doibase
  10.1038/nphoton.2013.100} {\bibfield  {journal} {\bibinfo  {journal} {Nat.
  Photon.}\ }\textbf {\bibinfo {volume} {7}},\ \bibinfo {pages} {517-520}
  (\bibinfo {year} {2013})}\BibitemShut {NoStop}%
\bibitem [{\citenamefont {Ockeloen}\ \emph {et~al.}(2013)\citenamefont
  {Ockeloen}, \citenamefont {Schmied}, \citenamefont {Riedel},\ and\
  \citenamefont {Treutlein}}]{Ockeloen2013}%
  \BibitemOpen
  \bibfield  {author} {\bibinfo {author} {\bibfnamefont {Caspar~F.}\
  \bibnamefont {Ockeloen}}, \bibinfo {author} {\bibfnamefont {Roman}\
  \bibnamefont {Schmied}}, \bibinfo {author} {\bibfnamefont {Max~F.}\
  \bibnamefont {Riedel}}, \ and\ \bibinfo {author} {\bibfnamefont {Philipp}\
  \bibnamefont {Treutlein}},\ }\bibfield  {title} {\enquote {\bibinfo {title}
  {Quantum metrology with a scanning probe atom interferometer},}\ }\href
  {\doibase 10.1103/PhysRevLett.111.143001} {\bibfield  {journal} {\bibinfo
  {journal} {Phys. Rev. Lett.}\ }\textbf {\bibinfo {volume} {111}},\ \bibinfo
  {pages} {143001} (\bibinfo {year} {2013})}\BibitemShut {NoStop}%
\bibitem [{\citenamefont {Budker}\ and\ \citenamefont
  {Romalis}(2007)}]{Budker2007}%
  \BibitemOpen
  \bibfield  {author} {\bibinfo {author} {\bibfnamefont {Dmitry}\ \bibnamefont
  {Budker}}\ and\ \bibinfo {author} {\bibfnamefont {Michael}\ \bibnamefont
  {Romalis}},\ }\bibfield  {title} {\enquote {\bibinfo {title} {Optical
  magnetometry},}\ }\href {\doibase 10.1038/nphys566} {\bibfield  {journal}
  {\bibinfo  {journal} {Nat. Phys.}\ }\textbf {\bibinfo {volume} {3}},\
  \bibinfo {pages} {227-234} (\bibinfo {year} {2007})}\BibitemShut {NoStop}%
\bibitem [{\citenamefont {Vasilakis}\ \emph {et~al.}(2011)\citenamefont
  {Vasilakis}, \citenamefont {Shah},\ and\ \citenamefont
  {Romalis}}]{Vasilakis2011a}%
  \BibitemOpen
  \bibfield  {author} {\bibinfo {author} {\bibfnamefont {G.}~\bibnamefont
  {Vasilakis}}, \bibinfo {author} {\bibfnamefont {V.}~\bibnamefont {Shah}}, \
  and\ \bibinfo {author} {\bibfnamefont {M.~V.}\ \bibnamefont {Romalis}},\
  }\bibfield  {title} {\enquote {\bibinfo {title} {Stroboscopic backaction
  evasion in a dense alkali-metal vapor},}\ }\href {\doibase
  10.1103/PhysRevLett.106.143601} {\bibfield  {journal} {\bibinfo  {journal}
  {Phys. Rev. Lett.}\ }\textbf {\bibinfo {volume} {106}},\ \bibinfo {pages}
  {143601} (\bibinfo {year} {2011})}\BibitemShut {NoStop}%
\bibitem [{\citenamefont {Behbood}\ \emph
  {et~al.}(2013{\natexlab{a}})\citenamefont {Behbood}, \citenamefont
  {Martin~Ciurana}, \citenamefont {Colangelo}, \citenamefont {Napolitano},
  \citenamefont {Mitchell},\ and\ \citenamefont {Sewell}}]{Behbood2013b}%
  \BibitemOpen
  \bibfield  {author} {\bibinfo {author} {\bibfnamefont {N.}~\bibnamefont
  {Behbood}}, \bibinfo {author} {\bibfnamefont {F.}~\bibnamefont
  {Martin~Ciurana}}, \bibinfo {author} {\bibfnamefont {G.}~\bibnamefont
  {Colangelo}}, \bibinfo {author} {\bibfnamefont {M.}~\bibnamefont
  {Napolitano}}, \bibinfo {author} {\bibfnamefont {M.~W.}\ \bibnamefont
  {Mitchell}}, \ and\ \bibinfo {author} {\bibfnamefont {R.~J.}\ \bibnamefont
  {Sewell}},\ }\bibfield  {title} {\enquote {\bibinfo {title} {Real-time vector
  field tracking with a cold-atom magnetometer},}\ }\href {\doibase
  http://dx.doi.org/10.1063/1.4803684} {\bibfield  {journal} {\bibinfo
  {journal} {Appl. Phys. Lett.}\ }\textbf {\bibinfo {volume} {102}},\ \bibinfo
  {eid} {173504} (\bibinfo {year} {2013}{\natexlab{a}})}\BibitemShut {NoStop}%
\bibitem [{\citenamefont {Sadler}\ \emph {et~al.}(2006)\citenamefont {Sadler},
  \citenamefont {Higbie}, \citenamefont {Leslie}, \citenamefont
  {Vengalattore},\ and\ \citenamefont {Stamper-Kurn}}]{Sadler2006}%
  \BibitemOpen
  \bibfield  {author} {\bibinfo {author} {\bibfnamefont {L.~E.}\ \bibnamefont
  {Sadler}}, \bibinfo {author} {\bibfnamefont {J.~M.}\ \bibnamefont {Higbie}},
  \bibinfo {author} {\bibfnamefont {S.~R.}\ \bibnamefont {Leslie}}, \bibinfo
  {author} {\bibfnamefont {M.}~\bibnamefont {Vengalattore}}, \ and\ \bibinfo
  {author} {\bibfnamefont {D.~M.}\ \bibnamefont {Stamper-Kurn}},\ }\bibfield
  {title} {\enquote {\bibinfo {title} {Spontaneous symmetry breaking in a
  quenched ferromagnetic spinor bose-einstein condensate},}\ }\href
  {http://dx.doi.org/10.1038/nature05094} {\bibfield  {journal} {\bibinfo
  {journal} {Nature}\ }\textbf {\bibinfo {volume} {443}},\ \bibinfo {pages}
  {312-315} (\bibinfo {year} {2006})}\BibitemShut {NoStop}%
\bibitem [{\citenamefont {Vengalattore}\ \emph {et~al.}(2007)\citenamefont
  {Vengalattore}, \citenamefont {Higbie}, \citenamefont {Leslie}, \citenamefont
  {Guzman}, \citenamefont {Sadler},\ and\ \citenamefont
  {Stamper-Kurn}}]{Vengalattore2007}%
  \BibitemOpen
  \bibfield  {author} {\bibinfo {author} {\bibfnamefont {M.}~\bibnamefont
  {Vengalattore}}, \bibinfo {author} {\bibfnamefont {J.~M.}\ \bibnamefont
  {Higbie}}, \bibinfo {author} {\bibfnamefont {S.~R.}\ \bibnamefont {Leslie}},
  \bibinfo {author} {\bibfnamefont {J.}~\bibnamefont {Guzman}}, \bibinfo
  {author} {\bibfnamefont {L.~E.}\ \bibnamefont {Sadler}}, \ and\ \bibinfo
  {author} {\bibfnamefont {D.~M.}\ \bibnamefont {Stamper-Kurn}},\ }\bibfield
  {title} {\enquote {\bibinfo {title} {High-resolution magnetometry with a
  spinor bose-einstein condensate},}\ }\href {\doibase
  10.1103/PhysRevLett.98.200801} {\bibfield  {journal} {\bibinfo  {journal}
  {Phys. Rev. Lett.}\ }\textbf {\bibinfo {volume} {98}},\ \bibinfo {pages}
  {200801} (\bibinfo {year} {2007})}\BibitemShut {NoStop}%
\bibitem [{\citenamefont {Liu}\ \emph {et~al.}(2009)\citenamefont {Liu},
  \citenamefont {Gomez}, \citenamefont {Maxwell}, \citenamefont {Turner},
  \citenamefont {Tiesinga},\ and\ \citenamefont {Lett}}]{Liu2009a}%
  \BibitemOpen
  \bibfield  {author} {\bibinfo {author} {\bibfnamefont {Y.}~\bibnamefont
  {Liu}}, \bibinfo {author} {\bibfnamefont {E.}~\bibnamefont {Gomez}}, \bibinfo
  {author} {\bibfnamefont {S.~E.}\ \bibnamefont {Maxwell}}, \bibinfo {author}
  {\bibfnamefont {L.~D.}\ \bibnamefont {Turner}}, \bibinfo {author}
  {\bibfnamefont {E.}~\bibnamefont {Tiesinga}}, \ and\ \bibinfo {author}
  {\bibfnamefont {P.~D.}\ \bibnamefont {Lett}},\ }\bibfield  {title} {\enquote
  {\bibinfo {title} {Number fluctuations and energy dissipation in sodium
  spinor condensates},}\ }\href {\doibase 10.1103/PhysRevLett.102.225301}
  {\bibfield  {journal} {\bibinfo  {journal} {Phys. Rev. Lett.}\ }\textbf
  {\bibinfo {volume} {102}},\ \bibinfo {pages} {225301} (\bibinfo {year}
  {2009})}\BibitemShut {NoStop}%
\bibitem [{\citenamefont {T{\'o}th}\ and\ \citenamefont
  {Mitchell}(2010)}]{Toth2010a}%
  \BibitemOpen
  \bibfield  {author} {\bibinfo {author} {\bibfnamefont {G{\'e}za}\
  \bibnamefont {T{\'o}th}}\ and\ \bibinfo {author} {\bibfnamefont {Morgan~W}\
  \bibnamefont {Mitchell}},\ }\bibfield  {title} {\enquote {\bibinfo {title}
  {Generation of macroscopic singlet states in atomic ensembles},}\ }\href
  {\doibase 10.1088/1367-2630/12/5/053007} {\bibfield  {journal} {\bibinfo
  {journal} {New J. Phys.}\ }\textbf {\bibinfo {volume} {12}},\ \bibinfo
  {pages} {053007} (\bibinfo {year} {2010})}\BibitemShut {NoStop}%
\bibitem [{\citenamefont {Hauke}\ \emph {et~al.}(2013)\citenamefont {Hauke},
  \citenamefont {Sewell}, \citenamefont {Mitchell},\ and\ \citenamefont
  {Lewenstein}}]{Hauke2013}%
  \BibitemOpen
  \bibfield  {author} {\bibinfo {author} {\bibfnamefont {P.}~\bibnamefont
  {Hauke}}, \bibinfo {author} {\bibfnamefont {R.~J.}\ \bibnamefont {Sewell}},
  \bibinfo {author} {\bibfnamefont {M.~W.}\ \bibnamefont {Mitchell}}, \ and\
  \bibinfo {author} {\bibfnamefont {M.}~\bibnamefont {Lewenstein}},\ }\bibfield
   {title} {\enquote {\bibinfo {title} {Quantum control of spin correlations in
  ultracold lattice gases},}\ }\href {\doibase 10.1103/PhysRevA.87.021601}
  {\bibfield  {journal} {\bibinfo  {journal} {Phys. Rev. A}\ }\textbf {\bibinfo
  {volume} {87}},\ \bibinfo {pages} {021601} (\bibinfo {year}
  {2013})}\BibitemShut {NoStop}%
\bibitem [{\citenamefont {Behbood}\ \emph
  {et~al.}(2013{\natexlab{b}})\citenamefont {Behbood}, \citenamefont
  {Colangelo}, \citenamefont {Martin~Ciurana}, \citenamefont {Napolitano},
  \citenamefont {Sewell},\ and\ \citenamefont {Mitchell}}]{Behbood2013a}%
  \BibitemOpen
  \bibfield  {author} {\bibinfo {author} {\bibfnamefont {N.}~\bibnamefont
  {Behbood}}, \bibinfo {author} {\bibfnamefont {G.}~\bibnamefont {Colangelo}},
  \bibinfo {author} {\bibfnamefont {F.}~\bibnamefont {Martin~Ciurana}},
  \bibinfo {author} {\bibfnamefont {M.}~\bibnamefont {Napolitano}}, \bibinfo
  {author} {\bibfnamefont {R.~J.}\ \bibnamefont {Sewell}}, \ and\ \bibinfo
  {author} {\bibfnamefont {M.~W.}\ \bibnamefont {Mitchell}},\ }\bibfield
  {title} {\enquote {\bibinfo {title} {Feedback cooling of an atomic spin
  ensemble},}\ }\href {\doibase 10.1103/PhysRevLett.111.103601} {\bibfield
  {journal} {\bibinfo  {journal} {Phys. Rev. Lett.}\ }\textbf {\bibinfo
  {volume} {111}},\ \bibinfo {pages} {103601} (\bibinfo {year}
  {2013}{\natexlab{b}})}\BibitemShut {NoStop}%
\bibitem [{\citenamefont {Puentes}\ \emph {et~al.}(2013)\citenamefont
  {Puentes}, \citenamefont {Colangelo}, \citenamefont {Sewell},\ and\
  \citenamefont {Mitchell}}]{Puentes2013a}%
  \BibitemOpen
  \bibfield  {author} {\bibinfo {author} {\bibfnamefont {G.}~\bibnamefont
  {Puentes}}, \bibinfo {author} {\bibfnamefont {G.}~\bibnamefont {Colangelo}},
  \bibinfo {author} {\bibfnamefont {R.~J.}\ \bibnamefont {Sewell}}, \ and\
  \bibinfo {author} {\bibfnamefont {M.~W.}\ \bibnamefont {Mitchell}},\
  }\bibfield  {title} {\enquote {\bibinfo {title} {Planar squeezing by quantum
  nondemolition measurement in cold atomic ensembles},}\ }\href {\doibase
  10.1088/1367-2630/15/10/103031} {\bibfield  {journal} {\bibinfo  {journal}
  {New J. Phys.}\ }\textbf {\bibinfo {volume} {15}},\ \bibinfo {pages} {103031}
  (\bibinfo {year} {2013})}\BibitemShut {NoStop}%
\bibitem [{\citenamefont {Eckert}\ \emph {et~al.}(2007)\citenamefont {Eckert},
  \citenamefont {Zawitkowski}, \citenamefont {Sanpera}, \citenamefont
  {Lewenstein},\ and\ \citenamefont {Polzik}}]{Eckert2007}%
  \BibitemOpen
  \bibfield  {author} {\bibinfo {author} {\bibfnamefont {K.}~\bibnamefont
  {Eckert}}, \bibinfo {author} {\bibfnamefont {\L{}.}\ \bibnamefont
  {Zawitkowski}}, \bibinfo {author} {\bibfnamefont {A.}~\bibnamefont
  {Sanpera}}, \bibinfo {author} {\bibfnamefont {M.}~\bibnamefont {Lewenstein}},
  \ and\ \bibinfo {author} {\bibfnamefont {E.~S.}\ \bibnamefont {Polzik}},\
  }\bibfield  {title} {\enquote {\bibinfo {title} {Quantum polarization
  spectroscopy of ultracold spinor gases},}\ }\href {\doibase
  10.1103/PhysRevLett.98.100404} {\bibfield  {journal} {\bibinfo  {journal}
  {Phys. Rev. Lett.}\ }\textbf {\bibinfo {volume} {98}},\ \bibinfo {pages}
  {100404} (\bibinfo {year} {2007})}\BibitemShut {NoStop}%
\bibitem [{\citenamefont {Eckert}\ \emph {et~al.}(2008)\citenamefont {Eckert},
  \citenamefont {Romero-Isart}, \citenamefont {Rodriguez}, \citenamefont
  {Lewenstein}, \citenamefont {Polzik},\ and\ \citenamefont
  {Sanpera}}]{Eckert2008}%
  \BibitemOpen
  \bibfield  {author} {\bibinfo {author} {\bibfnamefont {Kai}\ \bibnamefont
  {Eckert}}, \bibinfo {author} {\bibfnamefont {Oriol}\ \bibnamefont
  {Romero-Isart}}, \bibinfo {author} {\bibfnamefont {Mirta}\ \bibnamefont
  {Rodriguez}}, \bibinfo {author} {\bibfnamefont {Maciej}\ \bibnamefont
  {Lewenstein}}, \bibinfo {author} {\bibfnamefont {Eugene~S.}\ \bibnamefont
  {Polzik}}, \ and\ \bibinfo {author} {\bibfnamefont {Anna}\ \bibnamefont
  {Sanpera}},\ }\bibfield  {title} {\enquote {\bibinfo {title} {Quantum
  nondemolition detection of strongly correlated systems},}\ }\href {\doibase
  10.1038/nphys776} {\bibfield  {journal} {\bibinfo  {journal} {Nat. Phys.}\
  }\textbf {\bibinfo {volume} {4}},\ \bibinfo {pages} {50-54} (\bibinfo {year}
  {2008})}\BibitemShut {NoStop}%
\bibitem [{\citenamefont {Kubasik}\ \emph {et~al.}(2009)\citenamefont
  {Kubasik}, \citenamefont {Koschorreck}, \citenamefont {Napolitano},
  \citenamefont {de~Echaniz}, \citenamefont {Crepaz}, \citenamefont {Eschner},
  \citenamefont {Polzik},\ and\ \citenamefont {Mitchell}}]{Kubasik2009}%
  \BibitemOpen
  \bibfield  {author} {\bibinfo {author} {\bibfnamefont {M.}~\bibnamefont
  {Kubasik}}, \bibinfo {author} {\bibfnamefont {M.}~\bibnamefont
  {Koschorreck}}, \bibinfo {author} {\bibfnamefont {M.}~\bibnamefont
  {Napolitano}}, \bibinfo {author} {\bibfnamefont {S.~R.}\ \bibnamefont
  {de~Echaniz}}, \bibinfo {author} {\bibfnamefont {H.}~\bibnamefont {Crepaz}},
  \bibinfo {author} {\bibfnamefont {J.}~\bibnamefont {Eschner}}, \bibinfo
  {author} {\bibfnamefont {E.~S.}\ \bibnamefont {Polzik}}, \ and\ \bibinfo
  {author} {\bibfnamefont {M.~W.}\ \bibnamefont {Mitchell}},\ }\bibfield
  {title} {\enquote {\bibinfo {title} {Polarization-based light-atom quantum
  interface with an all-optical trap},}\ }\href {\doibase
  10.1103/PhysRevA.79.043815} {\bibfield  {journal} {\bibinfo  {journal} {Phys.
  Rev. A}\ }\textbf {\bibinfo {volume} {79}},\ \bibinfo {pages} {043815}
  (\bibinfo {year} {2009})}\BibitemShut {NoStop}%
\bibitem [{\citenamefont {de~Echaniz}\ \emph {et~al.}(2005)\citenamefont
  {de~Echaniz}, \citenamefont {Mitchell}, \citenamefont {Kubasik},
  \citenamefont {Koschorreck}, \citenamefont {Crepaz}, \citenamefont
  {Eschner},\ and\ \citenamefont {Polzik}}]{Echaniz2005}%
  \BibitemOpen
  \bibfield  {author} {\bibinfo {author} {\bibfnamefont {S.~R.}\ \bibnamefont
  {de~Echaniz}}, \bibinfo {author} {\bibfnamefont {M.~W.}\ \bibnamefont
  {Mitchell}}, \bibinfo {author} {\bibfnamefont {M.}~\bibnamefont {Kubasik}},
  \bibinfo {author} {\bibfnamefont {M.}~\bibnamefont {Koschorreck}}, \bibinfo
  {author} {\bibfnamefont {H.}~\bibnamefont {Crepaz}}, \bibinfo {author}
  {\bibfnamefont {J.}~\bibnamefont {Eschner}}, \ and\ \bibinfo {author}
  {\bibfnamefont {E.~S.}\ \bibnamefont {Polzik}},\ }\bibfield  {title}
  {\enquote {\bibinfo {title} {Conditions for spin squeezing in a cold 87 rb
  ensemble},}\ }\href {\doibase 10.1088/1464-4266/7/12/016} {\bibfield
  {journal} {\bibinfo  {journal} {J. Opt. B}\ }\textbf {\bibinfo {volume}
  {7}},\ \bibinfo {pages} {S548} (\bibinfo {year} {2005})}\BibitemShut
  {NoStop}%
\bibitem [{\citenamefont {Kuzmich}\ \emph {et~al.}(1998)\citenamefont
  {Kuzmich}, \citenamefont {Bigelow},\ and\ \citenamefont
  {Mandel}}]{Kuzmich1998}%
  \BibitemOpen
  \bibfield  {author} {\bibinfo {author} {\bibfnamefont {A.}~\bibnamefont
  {Kuzmich}}, \bibinfo {author} {\bibfnamefont {N.~P.}\ \bibnamefont
  {Bigelow}}, \ and\ \bibinfo {author} {\bibfnamefont {L.}~\bibnamefont
  {Mandel}},\ }\bibfield  {title} {\enquote {\bibinfo {title} {Atomic quantum
  nondemolition measurements and squeezing},}\ }\href
  {http://iopscience.iop.org/0295-5075/42/5/481} {\bibfield  {journal}
  {\bibinfo  {journal} {Europhys. Lett.}\ }\textbf {\bibinfo {volume} {42}},\
  \bibinfo {pages} {481} (\bibinfo {year} {1998})}\BibitemShut {NoStop}%
\bibitem [{\citenamefont {Wineland}\ \emph {et~al.}(1992)\citenamefont
  {Wineland}, \citenamefont {Bollinger}, \citenamefont {Itano}, \citenamefont
  {Moore},\ and\ \citenamefont {Heinzen}}]{Wineland1992}%
  \BibitemOpen
  \bibfield  {author} {\bibinfo {author} {\bibfnamefont {D.~J.}\ \bibnamefont
  {Wineland}}, \bibinfo {author} {\bibfnamefont {J.~J.}\ \bibnamefont
  {Bollinger}}, \bibinfo {author} {\bibfnamefont {W.~M.}\ \bibnamefont
  {Itano}}, \bibinfo {author} {\bibfnamefont {F.~L.}\ \bibnamefont {Moore}}, \
  and\ \bibinfo {author} {\bibfnamefont {D.~J.}\ \bibnamefont {Heinzen}},\
  }\bibfield  {title} {\enquote {\bibinfo {title} {Spin squeezing and reduced
  quantum noise in spectroscopy},}\ }\href {\doibase 10.1103/PhysRevA.46.R6797}
  {\bibfield  {journal} {\bibinfo  {journal} {Phys. Rev. A}\ }\textbf {\bibinfo
  {volume} {46}},\ \bibinfo {pages} {R6797-R6800} (\bibinfo {year}
  {1992})}\BibitemShut {NoStop}%
\bibitem [{\citenamefont {Colangelo}\ \emph {et~al.}(2013)\citenamefont
  {Colangelo}, \citenamefont {Sewell}, \citenamefont {Behbood}, \citenamefont
  {Ciurana}, \citenamefont {Triginer},\ and\ \citenamefont
  {Mitchell}}]{Colangelo2013a}%
  \BibitemOpen
  \bibfield  {author} {\bibinfo {author} {\bibfnamefont {G.}~\bibnamefont
  {Colangelo}}, \bibinfo {author} {\bibfnamefont {R.~J.}\ \bibnamefont
  {Sewell}}, \bibinfo {author} {\bibfnamefont {N.}~\bibnamefont {Behbood}},
  \bibinfo {author} {\bibfnamefont {F.~Martin}\ \bibnamefont {Ciurana}},
  \bibinfo {author} {\bibfnamefont {G.}~\bibnamefont {Triginer}}, \ and\
  \bibinfo {author} {\bibfnamefont {M.~W}\ \bibnamefont {Mitchell}},\
  }\bibfield  {title} {\enquote {\bibinfo {title} {Quantum atom-light
  interfaces in the gaussian description for spin-1 systems},}\ }\href
  {\doibase 10.1088/1367-2630/15/10/103007} {\bibfield  {journal} {\bibinfo
  {journal} {New J. Phys.}\ }\textbf {\bibinfo {volume} {15}},\ \bibinfo
  {pages} {103007} (\bibinfo {year} {2013})}\BibitemShut {NoStop}%
\bibitem [{\citenamefont {Madsen}\ and\ \citenamefont
  {M{\o}lmer}(2004)}]{MadsenPRA2004}%
  \BibitemOpen
  \bibfield  {author} {\bibinfo {author} {\bibfnamefont {L.~B.}\ \bibnamefont
  {Madsen}}\ and\ \bibinfo {author} {\bibfnamefont {K.}~\bibnamefont
  {M{\o}lmer}},\ }\bibfield  {title} {\enquote {\bibinfo {title} {Spin
  squeezing and precision probing with light and samples of atoms in the
  gaussian description},}\ }\href@noop {} {\bibfield  {journal} {\bibinfo
  {journal} {Phys. Rev. A}\ }\textbf {\bibinfo {volume} {70}},\ \bibinfo
  {pages} {052324} (\bibinfo {year} {2004})}\BibitemShut {NoStop}%
\bibitem [{\citenamefont {Pezz\'e}\ and\ \citenamefont
  {Smerzi}(2006)}]{Pezze2006a}%
  \BibitemOpen
  \bibfield  {author} {\bibinfo {author} {\bibfnamefont {Luca}\ \bibnamefont
  {Pezz\'e}}\ and\ \bibinfo {author} {\bibfnamefont {Augusto}\ \bibnamefont
  {Smerzi}},\ }\bibfield  {title} {\enquote {\bibinfo {title} {Phase
  sensitivity of a mach-zehnder interferometer},}\ }\href {\doibase
  10.1103/PhysRevA.73.011801} {\bibfield  {journal} {\bibinfo  {journal} {Phys.
  Rev. A}\ }\textbf {\bibinfo {volume} {73}},\ \bibinfo {pages} {011801}
  (\bibinfo {year} {2006})}\BibitemShut {NoStop}%
\end{thebibliography}

%

\end{document}